\documentclass[twocolumn,showpacs,preprintnumbers,amsmath,amssymb]{revtex4}
\usepackage{amssymb}
\usepackage{graphicx}
\usepackage{dcolumn}
\usepackage{bm}
\usepackage{textcomp}

\begin{document}
\title{ Three-intensity decoy state method for device independent quantum key distribution}
\author{Zong-Wen Yu$ ^{1,2}$, Yi-Heng Zhou$ ^{1}$,
and Xiang-Bin Wang$ ^{1,3\footnote{Email
Address: xbwang@mail.tsinghua.edu.cn}}$}

\affiliation{ \centerline{$^{1}$State Key Laboratory of Low
Dimensional Quantum Physics, Tsinghua University, Beijing 100084,
People¡¯s Republic of China}\centerline{$^{2}$Data Communication Science and Technology Research Institute, Beijing 100191, China}\centerline{$^{3}$ Shandong
Academy of Information and Communication Technology, Jinan 250101,
People¡¯s Republic of China}}

\begin{abstract}
We study the measurement device independent quantum key distribution (MDI-QKD) in practice with limited resource, when there are only 3 different states in implementing the decoy-state method. We present a more tightened explicit formula to estimate the lower bound of the yield of two-single-photon pulses. Moreover, we show that the bounding of this yield and phase flip error of single photon pulse pairs can be further improved by using other constraints which can be solved by a simple and explicit program. Results of numerical simulation for key rates with both the improved explicit formula and the program are presented. It shows that the results obtained with our methods here can significantly improve the key rate and secure distance of MDI QKD with only three intensities.
\end{abstract}


\maketitle


\section{Introduction}
Security for real set-ups of quantum key distribution (QKD)~\cite{BB84,GRTZ02} has become a major problem in the area in the recent years. The major problems here include the imperfection of source and the limited efficiency of the detection device. The decoy state method~\cite{ILM,H03,wang05,LMC05,AYKI,haya,peng,wangyang,rep,njp} can help to make a set-up with an imperfect single photon source be as secure as that with a perfect single photon
source~\cite{PNS1,PNS}.

Besides the source imperfection, the limited detection is another threaten to the security~\cite{lyderson}.
Theories of the device independent security proof~\cite{ind1} have been proposed
to overcome the problem. However, these theories cannot apply to the existing real set-ups because violation of
Bell's inequality cannot be strictly demonstrated by existing set-ups.

Very recently, an idea of measurement device independent QKD (MDI-QKD) was proposed based on the idea of
entanglement swapping~\cite{ind3,ind2}.
There, one can make secure QKD
simply by virtual entanglement swapping, i.e., Both Alice and Bob sends BB84 states to the relay which can be controlled by un-trusted third party (UTP). After the UTP announced his measurement outcome, Alice and Bob will post select those bits  corresponding to a successful event and prepared in the same basis for further processing. In the realization, Alice and Bob can really use entanglement pairs~\cite{ind3} and measure halves of the pair inside the lab before sending another halves to the UTP. In this way, the decoy-state method is not necessary even though imperfect entangled pairs (such as the states generated by the type II parametric down conversion) are used. Even though there are multi-pair events with small probability, these events do not affect the security. Alice and Bob only need to check the error rates of their post selected bits. However, in our existing technologies, high quality entangled-pair-state generation  can not be done efficiently. In the most matured technology, the generation rate is lower than 1 from 1000 pump pulses. If we want to obtain a higher key rate, we can choose to directly use an imperfect single-photon source such as the coherent state~\cite{ind2}. If we choose this, we must implement the decoy-state method for security. This has been discussed in Ref.~\cite{ind2}, and calculation formulas
for the practical decoy-state implementation with only a few different states was first presented in \cite{wangPRA2013}, and then further studied  both experimentally~\cite{tittel1,tittel2,liuyang} and theoretically\cite{qing3,LiangPRA2013,lopa,han,curty}.
In particular, Tittel's group\cite{tittel1,tittel2} did the MDI QKD experiment with  3 intensities\cite{wangPRA2013}, in the laboratory over more than 80 km of spooled fiber, as well as across different locations within the city of Calgary. By developing up-conversion single-photon detectors with high efficiency and low noise, Liu et al did it over a 50-km fiber link\cite{liuyang} and transmitted a 24192-image with one-time pad protocol.
The pioneering experiments in Calgary\cite{tittel1} and in Shanghai\cite{liuyang} make a big step towards the final goal of real application because they clearly show the practical feasibility of MDI QKD. Sun et al\cite{LiangPRA2013} presented a variant formula of 3-intensity MDI QKD with numerical simulation. However, the earlier formula\cite{wangPRA2013} actually behaves better than Sun et al result. Xu et al\cite{lopa} studied the more general case when each sides use 3 non-vacuum states. One can see that the major formula there is identical with the one in Ref.\cite{wangPRA2013} in the case when the weakest pulse is vacuum and both sides use the same 3 intensities. Qin and Wang\cite{qing3} and Zhou et al\cite{han} studied the MDI QKD with heralded single-photon sources. Curty et al studied some finite-key effects\cite{curtty}.

There are two directions for the future study of the MDI QKD. One is to improve the experimental techniques, so as to improve the robustness and efficiency. The other is to upgrade the theoretical results so as to obtain a higher key rate given the same experimental data.

Here in this work, we shall first give better explicit formulas of 3-state decoy-state method for the MDI-QKD . We then estimate the infimum of yield and the supremum of error rate  single-photon pulse pairs to the UTP with a simple and efficient program. In the fifth section, we present the numerical simulations. The article is ended with a concluding remark.

\section{Decoy-state method with only 3 states for MDI-QKD}
In the protocol, each time a pulse-pair (two-pulse state) is sent to the relay for detection. The relay is controlled by an UTP. The UTP will announce whether the pulse-pair has caused a successful event.
Those bits corresponding to successful events will be post-selected and further processed for the final key. Since real set-ups only use imperfect single-photon sources, we need the decoy-state method for security.

We assume Alice (Bob) has three sources, $o_A,x_A,y_A$ ($o_B,x_B,y_B$) which can only emit three different states $\rho_{o_A}=|0\rangle\langle 0|, \rho_{x_A}, \rho_{y_A}$ ($\rho_{o_B}=|0\rangle\langle 0|, \rho_{x_B}, \rho_{y_B}$), respectively, in photon number space.
Suppose
\begin{eqnarray}
\rho_{x_A}=\sum_{k} a_k |k\rangle\langle k|,&\quad& \rho_{y_A}=\sum_{k} a_k' |k\rangle\langle k|;\\
\rho_{x_B}=\sum_{k} b_k |k\rangle\langle k|,&\quad& \rho_{y_B}=\sum_{k} b_k' |k\rangle\langle k|,
\end{eqnarray}
and we request the states satisfy the following very important condition:
\begin{equation}\label{cond1}
\frac{a_k'}{a_k}\ge \frac{a_2'}{a_2}\ge \frac{a_1'}{a_1};\quad \frac{b_k'}{b_k}\ge \frac{b_2'}{b_2}\ge \frac{b_1'}{b_1},
\end{equation}
for $k\ge 2$.
The imperfect sources used in practice such as the coherent state source, the heralded source out of the parametric-down conversion,  satisfy the above restriction. Given a specific type of source, the above listed different states have different averaged photon numbers (intensities), therefore the states can be obtained by controlling the light intensities.
At each time, Alice will randomly select one of her 3 sources to emit a pulse, and so does Bob. The pulse form Alice and the pulse from Bob form a pulse pair and are sent to the un-trusted relay.  We regard equivalently that each time a two-pulse source is selected and a pulse pair (one pulse from Alice, one pulse from Bob) is emitted.
There are many different two-pulse sources used in the protocol. We denote $\alpha\beta$ for the two pulse source when the pulse-pair is produced by source
$\alpha$ at Alice's side and source $\beta$ at Bob's side, $\alpha$ can be one of $\{o_A,x_A,y_A\}$ and $\beta$
can be one of $\{o_B,x_B,y_B\}$. For example, at a certain time $j$ Alice uses source $o_A$ and Bob uses source $y_B$, we say the pulse pair is emitted by source $o_Ay_B$.

In the protocol, two different bases, $Z$ basis consisting of horizontal polarization $|H\rangle\langle H|$  and vertical polarization $|V\rangle\langle V|$, and $X$ basis consisting of $\pi/4$ and $3\pi/4$ polarizations are used. The density operator in photon number space alone does not describe the state in the composite space. We shall apply the the decoy-state method analysis in the same basis (e.g., $Z$ basis or $X$ basis) for pulses from sources $x_A,x_B,y_A,y_B$. Therefore we only need consider the density operators in the photon number space. For simplicity, we consider pulses from source prepared in $Z$ basis first.

According to the decoy-state theory, the yield of a certain set of pulse pairs is defined as the  happening rate of a successful event (announced by the UTP) corresponding to pulse pairs out of the set. Mathematically, the yield is $n/N$ where $n$ is the number of successful events happened corresponding to pulse pairs from the set and $N$ is the number of pulse pairs in the set. Obviously, if we regard the pulse pairs of two-pulse source $\alpha\beta$ as a set, the yield $S_{\alpha\beta}$ for source $\alpha\beta$ is $S_{\alpha\beta}= \frac{n_{\alpha\beta}}{N_{{\alpha\beta}}}$, where $n_{\alpha\beta}$ is the number of successful events happened corresponding to pulse pairs from source $\alpha\beta$ and $N_{\alpha\beta}$ is the number of times source ${\alpha\beta}$ are used. In the protocol, there are 9 different two-pulse sources. The yields of these 9 sources can be directly calculated from the observed experimental data $n_{\alpha\beta}$ and $N_{\alpha\beta}$. We use capital letter $S_{\alpha\beta}$ for these {\em known} values.

We can regard any source as a composite source that consists of many (virtual) sub-sources, if the source state can be be written in a convex form of different density operators. For example, two-pulse source $y_Ay_B$ includes a sub-source of pulse pairs of state $\rho_1\otimes\rho_1$ ($\rho_1=|1\rangle\langle 1|$) with weight $a_1'b_1'$. This is to say, after we have used source $y_Ay_B$ for $N$ times, we have actually used sub-source of state $\rho_1\otimes \rho_1$ for $a_1'b_1' N$ times, asymptotically. Similarly, the source $x_Ax_B$ also includes a sub-source of state $\rho_1\otimes \rho_1$ with weight $a_1b_1$. These two sub-sources of state $\rho_1\otimes \rho_1$ must have the same yield $s_{11}$ because they have the same two-pulse state and the pulse pairs are randomly mixed. Most generally, denote $s,s'$ as the yields of two sets of pulses, if pulse pairs of these two sets are randomly mixed and all pulses have the same density operator, then
\begin{equation}\label{ele1}
 s=s'
\end{equation}
asymptotically. This is the elementary assumption of the decoy-state theory.

In the protocol, since each sources are randomly chosen, pulses from each sub-sources or sources are also randomly mixed. Therefore, the yield of a sub-source or a source is dependent on the {\em state} only, it is independent of which physical source the pulses are from. Therefore, we can also define the yield of a certain state: whenever a pulse pair of that state is emitted, the probability that a successful event happens.
Denote
\begin{equation}
 \Omega_{\alpha\beta} = \rho_\alpha \otimes  \rho_\beta
\end{equation} for a two-pulse state. The yield of such a state is also the yield of any source which produces state $\Omega_{\alpha\beta}$ only, or the yield of a sub-source from {\em any} source, provided that the state of the pulse pairs of the sub-source is $\Omega_{\alpha\beta}$. Note that, we don't always know the value of yield of a state. Because we don't know which sub-source was used at which time.
We shall use the lower case symbol $s_{\alpha,\beta}$ to denote the yield of  state $\Omega_{\alpha,\beta}$.
In general, the yields of a sub-source (a state), such as $s_{11}$  is not directly known from the experimental data. But some of them can be deduced from the yields of different real sources.  Define $\rho_0 =|0\rangle\langle 0|$. According to Eq.(\ref{ele1}), if $\alpha\in\{0,x_A,y_A\}$ and $\beta\in\{0,x_B,y_B\}$, we have
\begin{equation}\label{sub}
  s_{\alpha\beta} = S_{\tilde\alpha\tilde\beta}
\end{equation}
with the mapping of $\tilde \alpha = (o_A,x_A,y_A)$ for $\alpha = (0,x_A,y_A)$, respectively; and $\tilde \beta = (o_B,x_B,y_B)$ for $\beta = (0,x_B, y_B)$, respectively.
To understand the meaning of the equation above, we take an example for pulses from source $y_Ay_B$. By writing the state of this source in the convex form we immediately know that it includes a sub-source of state $\rho_0\otimes \rho_{y_B}$. By observing the results caused by source $y_Ay_B$ itself we have no way to know the yield of this sub-source because we don't know exactly which time source $y_A$ emits a vacuum pulse when we use it. However, the state of this sub-source is the same with the state of the real source $o_Ay_B$, therefore the yield of any sub-source of state $\rho_0\otimes\rho_{y_B}$ must be just the yield of the real source $o_Ay_B$, which can be directly observed in the experiment. Mathematically, this is $s_{0y_B}=S_{o_Ay_B}$, where  the right hand side is the known value of yield of real source $o_Ay_B$, the left hand side is the yield of a virtual sub-source from real source $y_Ay_B$.

Our first major task is to deduce $s_{11}$ from the known values, i.e., to formulate $s_{11}$, the yield of state $|1\rangle\langle 1|\otimes |1\rangle\langle 1|$ in capital-letter symbols $\{S_{\alpha\beta}\}$.
We shall use the following convex proposition to do the calculation.

Denote  $S$ to be the yield of a certain source of state $\Omega$. If $\Omega$ has the convex forms of $\Omega = \sum_{\alpha\beta} c_{\alpha\beta} \Omega_{\alpha\beta}$, we have
\begin{equation}\label{convex}
  S=\sum_{\alpha,\beta}c_{\alpha\beta} s_{\alpha\beta}.
\end{equation}
This equation is simply the fact that the total number of successful events caused by pulses from a certain set  is equal to the summation of the numbers of successful events caused by pulses from each  sub-sets.

Consider the convex forms of source $x_Ax_B$, $x_Ay_B$, $y_Ax_B$ and source $y_Ay_B$. Without causing any ambiguity, we omit the subscripts $A$ and $B$ in the following of this paper. Explicitly,
\begin{eqnarray}
  \tilde{S}_{xx}= a_1 b_1 s_{11}+a_1 b_2 s_{12} +a_2 b_1 s_{21} +a_2 b_2 s_{22}+J_{xx},&& \label{xx} \\
  \tilde{S}_{xy}= a_1 b'_1 s_{11}+a_1 b'_2 s_{12} +a_2 b'_1 s_{21} +a_2 b'_2 s_{22}+J_{xy},&& \label{xy} \\
  \tilde{S}_{yx}= a'_1 b_1 s_{11}+a'_1 b_2 s_{12} +a'_2 b_1 s_{21} +a'_2 b_2 s_{22}+J_{yx},&& \label{yx} \\
  \tilde{S}_{yy}= a'_1 b'_1 s_{11}+a'_1 b'_2 s_{12} +a'_2 b'_1 s_{21} +a'_2 b'_2 s_{22}+J_{yy},&& \label{yy}
\end{eqnarray}
where
\begin{eqnarray}
  \tilde{S}_{xx}&=& S_{xx}-a_0 S_{0x}-b_0 S_{x0}+a_0 b_0 S_{00}, \label{tSxx} \\
  \tilde{S}_{xy}&=& S_{xy}-a_0 S_{0y}-b'_0 S_{x0}+a_0 b'_0 S_{00}, \label{tSxy} \\
  \tilde{S}_{yx}&=& S_{yx}-a'_0 S_{0x}-b_0 S_{y0}+a'_0 b_0 S_{00}, \label{tSyx} \\
  \tilde{S}_{yy}&=& S_{yy}-a'_0 S_{0y}-b'_0 S_{y0}+a'_0 b'_0 S_{00}, \label{tSyy}
\end{eqnarray}
and
\begin{eqnarray}
  J_{xx}= \sum_{(m,n)\in J_0}a_m b_n s_{mn}, & &
  J_{xy}= \sum_{(m,n)\in J_0}a_m b'_n s_{mn}, \nonumber \\
  J_{yx}= \sum_{(m,n)\in J_0}a'_m b_n s_{mn},& &
  J_{yy}= \sum_{(m,n)\in J_0}a'_m b'_n s_{mn}, \nonumber
\end{eqnarray}
with $J_0=\{(m,n)|m\geq 1,n\geq 1,m+n\geq 4,(m,n)\neq(2,2)\}$.

In order to get a lower bound of $s_{11}$, we should derive the expression of $s_{11}$ with Eqs.(\ref{xx}-\ref{yy}) firstly. Combining Eqs.(\ref{xx}-\ref{yx}), we obtain the expression of $s_{11}$ by eliminating $s_{12}$ and $s_{21}$ such that
\begin{equation}\label{s11with123}
  s_{11}=s_{11}^{(123)}+\sum_{(m,n)\in J_{1}}f_{11}^{(123)}(m,n)s_{mn},
\end{equation}
where $J_{1}=\{(m,n)|m\geq 1, n\geq 1,m+n\geq 4\}$,
\begin{widetext}
\begin{equation}\label{s11of123}
  s_{11}^{(123)}=\frac{(a_1 a'_2 b_1 b'_2-a'_1 a_2 b'_1 b_2)\tilde{S}_{xx}-b_1 b_2(a_1 a'_2-a'_1 a_2)\tilde{S}_{xy}-a_1 a_2 (b_1 b'_2-b'_1 b_2)\tilde{S}_{yx}}{a_1 b_1(a_1 a'_2-a'_1 a_2)(b_1 b'_2-b'_1 b_2)},
\end{equation}
and
\begin{equation}\label{f11of123}
  f_{11}^{(123)}(m,n)=\frac{a_2 b_n(a_1 a'_m-a'_1 a_m)(b_1 b'_2-b'_1 b_2)+a_m b_1(a_1 a'_2-a'_1 a_2)(b_2 b'_n-b'_2 b_n)}{a_1 b_1(a_1 a'_2-a'_1 a_2)(b_1 b'_2-b'_1 b_2)}.
\end{equation}
\end{widetext}
In these expressions, we use the superscript $*^{(123)}$ to denote the result obtained with the first three equations from Eqs.(\ref{xx}-\ref{yy}). Under the conditions presented in Eq.(\ref{cond1}), we can easily find out that $(a_1 a'_2-a'_1 a_2)\geq 0$, $(b_1 b'_2-b'_1 b_2)\geq 0$, $(a_1 a'_m-a'_1 a_m)\geq 0$ for all $m\geq 1$ and $(b_2 b'_n-b'_2 b_n)\geq 0$ for all $n\geq 2$. Then we know that $f_{11}^{(123)}(m,n)\geq 0$ hold for all $(m,n)\in J_1$. With this fact, we obtain a lower bound from Eq.(\ref{s11with123}) by setting $s_{mn}=0,(m,n)\in J_1$ such that
\begin{equation}\label{s11L123}
  \underline{s_{11}}=s_{11}^{(123)}\leq s_{11},
\end{equation}
where $s_{11}^{(123)}$ is defined by Eq.(\ref{s11of123}). This and Eq.(\ref{s11of123}) are our major formulas for the decoy-state method implementation for MDI-QKD in this section.

Similarly, we can get other expressions with choosing any other three equations from Eqs.(\ref{xx}-\ref{yy}). For example, we choose Eqs.(\ref{xx}-\ref{xy},\ref{yy}). By eliminating $s_{12}$ and $s_{21}$, we get another expression of $s_{11}$ such that
\begin{equation}\label{s11with124}
  s_{11}= s_{11}^{(124)}+\sum_{(m,n)\in J_1} f_{11}^{(124)}(m,n)s_{mn},
\end{equation}
where
\begin{widetext}
\begin{equation}\label{s11of124}
  s_{11}^{(124)}=\frac{b'_1 b'_2(a_1 a'_2-a'_1 a_2)\tilde{S}_{xx}+(a'_1 a_2 b_1 b'_2-a_1 a'_2 b'_1 b_2)\tilde{S}_{xy}-a_1 a_2 (b_1 b'_2-b'_1 b_2) \tilde{S}_{yy}}{a_1 b'_1(a_1 a'_2-a'_1 a_2)(b_1 b'_2-b'_1 b_2)},
\end{equation}
and
\begin{equation}\label{f11of124}
  f_{11}^{(124)}(m,n)=\frac{a_2 b'_n(a_1 a'_m-a'_1 a_m)(b_1 b'_2-b'_1 b_2)+a_m b'_1(a_1 a'_2-a'_1 a_2)(b_2 b'_n-b'_2 b_n)}{a_1 b'_1(a_1 a'_2-a'_1 a_2)(b_1 b'_2-b'_1 b_2)}.
\end{equation}
\end{widetext}
Under the conditions presented in Eq.(\ref{cond1}), we can also find out that $f_{11}^{(124)}(m,n)\geq 0$ for all $(m,n)\in J_1$. Then we know that $s_{11}^{(124)}$ is alos a lower bound of $s_{11}$. On the other hand, by comparing $f_{11}^{(123)}(m,n)$ and $f_{11}^{(124)}(m,n)$, we have
\begin{eqnarray}\label{dif123and124}
  & &f_{11}^{(123)}(m,n)-f_{11}^{(124)}(m,n) \nonumber \\
  &=&-\frac{a_2(a_1 a'_m-a'_1 a_m)(b_1 b'_n-b'_1 b_n)}{a_1 b_1 b'_1 (a_1 a'_2 -a'_1 a_2)}\leq 0,
\end{eqnarray}
for any $(m,n)\in J_1$. Then we know that
\begin{equation}\label{s11of123and124}
  s_{11}^{(123)}\geq s_{11}^{(124)},
\end{equation}
with Eq.(\ref{s11with123}) and Eq.(\ref{s11with124}). With the relation presented in Eq.(\ref{s11of123and124}) we know that the lower bound $s_{11}^{(123)}$ is tighter than the lower bound $s_{11}^{(124)}$. In the same way, we can get another two lower bounds $s_{11}^{(134)}$ and $s_{11}^{(234)}$ of $s_{11}$ with Eqs.(\ref{xx},\ref{yx}-\ref{yy}) and Eqs.(\ref{xy}-\ref{yy}) respectively. Furthermore, we can also prove that
\begin{equation}\label{dif34}
  s_{11}^{(123)}\geq s_{11}^{(134)}, \quad s_{11}^{(123)}\geq s_{11}^{(234)}.
\end{equation}

Now we only consider Eq.(\ref{xx}) and Eq.(\ref{yy}). By eliminating $s_{12}$ or $s_{21}$ respectively, we get two expressions of $s_{11}$ such that
\begin{eqnarray}
  s_{11}&=&s_{11}^{(14a)}+\sum_{(m,n)\in J_{2}} f_{11}^{(14a)}(m,n) s_{mn}, \label{s11Ka} \\
  s_{11}&=&s_{11}^{(14b)}+\sum_{(m,n)\in J_{2}} f_{11}^{(14a)}(m,n) s_{mn}, \label{s11Kb}
\end{eqnarray}
where $J_2=\{(m,n)|m\geq 1,n\geq 1, m+n\geq 3\}$,
\begin{eqnarray}
  s_{11}^{(14a)}&=&\frac{a'_1 b'_2 \tilde{S}_{xx}-a_1 b_2 \tilde{S}_{yy}}{a_1 a'_1(b_1 b'_2-b'_1 b_2)}, \label{s11of14a} \\
  s_{11}^{(14b)}&=&\frac{a'_2 b'_1 \tilde{S}_{xx}-a_2 b_1 \tilde{S}_{yy}}{b_1 b'_1(a_1 a'_2-a'_1 a_2)}, \label{s11of14a}
\end{eqnarray}
and
\begin{eqnarray}
  f_{11}^{(14a)}(m,n)&=&\frac{a_1 b_2 a'_m b'_n-a'_1 b'_2 a_m b_n}{a_1 a'_1(b_1 b'_2-b'_1 b_2)}, \label{f11of14a} \\
  f_{11}^{(14b)}(m,n)&=&\frac{a_2 b_1 a'_m b'_n-a'_2 b'_1 a_m b_n}{b_1 b'_1(a_1 a'_2-a'_1 a_2)}. \label{f11of14a}
\end{eqnarray}
For any sources used in the protocol, we must have either $K_a=\frac{a'_1 b'_2}{a_1 b_2}\leq \frac{a'_2 b'_1}{a_2 b_1}=K_b$ or $K_a\geq K_b$. Suppose the former one holds, we can easily find out that $f_{11}^{(14a)}(m,n)\geq 0$ for all $(m,n)\in J_2$ and $s_{11}^{(14a)}$ is a lower bound of $s_{11}$. On the other hand, if $K_a\geq K_b$ holds, we have $f_{11}^{(14b)}(m,n)\geq 0$ for all $(m,n)\in J_2$ and $s_{11}^{(14b)}$ is a lower bound of $s_{11}$. Considering the following two relations
\begin{equation}\label{kaKb}
  K_a-K_b =\frac{a'_1 a_2 b_1 b'_2-a_1 a'_2 b'_1 b_2}{a_1 a_2 b_1 b_2},
\end{equation}
and
\begin{eqnarray}\label{f11of14ab}
  & &f_{11}^{(14a)}(m,n)-f_{11}^{(14b)}(m,n) \nonumber \\
  &=&\frac{(a_1 a'_m b_1 b'_n-a'_1 a_m b'_1 b_n)(a'_1 a_2 b_1 b'_2-a_1 a'_2 b'_1 b_2)}{a_1 a'_1 b_1 b'_1 (a_1 a'_2-a'_1 a_2)(b_1 b'_2-b'_1 b_2)},
\end{eqnarray}
we know that $K_a-K_b$ and $f_{11}^{(14a)}-f_{11}^{(14b)}$ have the same sign which means that they are both positive or negative simultaneously. Then we can write the lower bound of $s_{11}$ with Eq.(\ref{xx}) and Eq.(\ref{yy}) into the following compact form
\begin{equation}\label{s11of14}
  s_{11}^{(14)}=\min\{s_{11}^{(14a)},s_{11}^{(14b)}\},
\end{equation}
that is the result presented in Ref.~\cite{wangPRA2013}. In the coming, we will prove that the lower bound $s_{11}^{(123)}$ given in Eq.(\ref{s11of123}) is more tightly than $s_{11}^{(14)}$. Firstly, if we suppose $K_a\leq K_b$ holds, then we know that $a'_1 a_2 b_1 b'_2\leq a_1 a'_2 b'_1 b_2$ and $s_{11}^{(14)}=s_{11}^{(14a)}$. For any $(m,n)\in J_1$ we have
\begin{eqnarray}
  & &f_{11}^{(123)}(m,n)-f_{11}^{(14a)}(m,n) \nonumber \\
  &=&-\frac{(a_1a'_m-a'_1 a_m)D_a}{a_1 a'_1 b_1 (a_1 a'_2 -a'_1 a_2)(b_1 b'_2 -b'_1 b_2)},
\end{eqnarray}
where $D_a=(a_1 a'_2 b_1 b_2 b'_n+a'_1 a_2 b'_1 b_2 b_n-a'_1 a_2 b_1 b'_2 b_n-a'_1 a_2 b_1 b_2 b'_n)\geq b_2 (a_1 a'_2-a'_1 a_2)(b_1 b'_n-b'_1 b_n)$. Then we know that
\begin{eqnarray}
  & &f_{11}^{(123)}(m,n)-f_{11}^{(14a)}(m,n) \nonumber \\
  &\leq & -\frac{b_2 (a_1 a'_m-a'_1 a_m)(b_1 b'_n-b'_1 b_n)}{a_1 a'_1 b_1 (b_1 b'_2-b'_1 b_2)} \leq 0.
\end{eqnarray}
We can easily know that $s_{11}^{(123)}\geq s_{11}^{(14)}$ when $K_a\leq K_b$ with this equation. Secondly, if we suppose $K_a\geq K_b$ holds, we can easily prove that $f_{11}^{(123)}(m,n)-f_{11}^{(14b)}(m,n)\leq 0$ for all $(m,n)\in J_1$ within the same way. Then we get $s_{11}^{(123)}\geq s_{11}^{(14)}$ when $K_a\geq K_b$.

In the last part of this section, we will derive another lower bound of $s_{11}$ with those four Eqs.(\ref{xx}-\ref{yy}). The idea presented in Refs.~\cite{wangPRA2013,LiangPRA2013} inspire us to do the following deduction
\begin{widetext}
\begin{eqnarray}\label{s11ofalphaS}
  \tilde{S}_{yy}-\tilde{S}_{xx}&=& (a'_1 b'_1-a_1 b_1) s_{11}+\sum_{n\geq 2}(a'_1 b'_n-a_1 b_n)s_{1n} +\sum_{m\geq 2}(a'_m b'_1-a_m b_1) s_{m1}+\sum_{m,n\geq 2}(a'_m b'_n-a_m b_n)s_{mn} \nonumber \\
  &\geq& (a'_1 b'_1-a_1 b_1) s_{11}+A\sum_{n\geq 2}(a'_1 b_n+a_1 b'_n)s_{1n}+B\sum_{m\geq 2}(a'_m b_1 +a_m b'_1) s_{m1}+C\sum_{m,n\geq 2}(a'_m b_n +a_m b'_n)s_{mn} \nonumber \\
  &\geq& (a'_1 b'_1-a_1 b_1) s_{11}+\alpha\left[\sum_{n\geq 2}(a'_1 b_n+a_1 b'_n)s_{1n}+\sum_{m\geq 2}(a'_m b_1 +a_m b'_1) s_{m1}+\sum_{m,n\geq 2}(a'_m b_n +a_m b'_n)s_{mn}\right] \nonumber \\
  &=& (a'_1 b'_1-a_1 b_1) s_{11}+\alpha \left( \tilde{S}_{xy}-a_1 b'_1 s_{11}+\tilde{S}_{yx}-a'_1 b_1 s_{11}\right) \nonumber \\
  &=&\left[ a'_1 b'_1-a_1 b_1 -\alpha(a_1 b'_1+a'_1 b_1)\right]s_{11}+\alpha(\tilde{S}_{xy}+\tilde{S}_{yx}),
\end{eqnarray}
\end{widetext}
where we have used the condition presented in Eq.(\ref{cond1}), and $\alpha=\min\{A,B,C\}$ with
\begin{equation*}
  A=\frac{a'_1 b'_2-a_1 b_2}{a'_1 b_2+a_1 b'_2}, \, B=\frac{a'_2 b'_1 -a_2 b_1}{a'_2 b_1 +a_2 b'_1}, \, C=\frac{a'_2 b'_2-a_2 b_2}{a'_2 b_2 +a_2 b'_2}.
\end{equation*}
Actually, under the condition in Eq.(\ref{cond1}), we know that
\begin{eqnarray*}
  A-C&=&-\frac{(b_2^2+{b'_2}^2)(a_1 a'_2-a'_1 a_2)}{(a'_1 b_2+a_1 b'_2)(a'_2 b_2+a_2 b'_2)}\leq 0, \\
  B-C&=&-\frac{(a_2^2+{a'_2}^2)(b_1 b'_2-b'_1 b_2)}{(a'_2 b_1+a_2 b'_1)(a'_2 b_2+a_2 b'_2)}\leq 0.
\end{eqnarray*}
Then $\alpha$ can be written as $\alpha=\min\{A,B\}$. According to the relation presented in Eq.(\ref{s11ofalphaS}), we obtain the other expression of $s_{11}$
\begin{equation}\label{s11withalpha}
  s_{11}=s_{11}^{(\alpha)}+\sum_{(m,n)\in J_{2}}f_{11}^{(\alpha)}(m,n)s_{mn},
\end{equation}
where
\begin{equation}\label{s11ofalpha}
  s_{11}^{(\alpha)}=\frac{\tilde{S}_{xx}-\tilde{S}_{yy}+\alpha(\tilde{S}_{xy}+\tilde{S}_{yx})}{a_1 b_1 -a'_1 b'_1+\alpha(a_1 b'_1+a'_1 b_1)},
\end{equation}
and
\begin{equation}\label{f11ofalpha}
  f_{11}^{(\alpha)}(m,n)=-\frac{a_m b_n-a'_m a'_n +\alpha(a_m b'_n+a'_m b_n)}{a_1 b_1 -a'_1 b'_1 +\alpha(a_1 b'_1 +a'_1 b_1)}.
\end{equation}
With the condition presented in Eq.(\ref{cond1}), we can easily prove that $f_{11}^{(\alpha)}(m,n)\geq 0$ for all $(m,n)\in J_2$. So we know that $s_{11}^{(\alpha)}$ is the other lower bound of $s_{11}$.
In the coming, we will discuss the relation among $s_{11}^{(\alpha)}$, $s_{11}^{(14)}$ and $s_{11}^{(123)}$.

Firstly, we consider the special case with $a_k=b_k$ and $a'_k=b'_k$ for any $k\geq 1$. In this case, we have $K_a=K_b=\frac{a'_1 a'_2}{a_1 a_2}$, $A=B=\frac{a'_1 a'_2-a_1 a_2}{a'_1 a_2+a_1 a'_2}$ and
\begin{eqnarray*}
  & &f_{11}^{(\alpha)}(m,n)-f_{11}^{(14)}(m,n) \nonumber \\
  &=&\frac{(a'_1 a'_2 -a_1 a_2)(a_1 a'_m -a'_1 a_m)(a_1 a'_n-a'_1 a_n)}{a_1 a'_1 (a_1^2+{a'_1}^2)(a_1 a'_2-a'_1 a_2)}\geq 0,
\end{eqnarray*}
for any $(m,n)\in J_2$. Then we know that $s_{11}^{(14)}\geq s_{11}^{(\alpha)}$ in this case.

Secondly, for the general case, we can prove that
\begin{equation}\label{s11of123withalpha}
  s_{11}^{(123)}\geq s_{11}^{(\alpha)}.
\end{equation}
In the situation with $A\leq B$, we have
\begin{equation*}
  A-B=\frac{D_{A_1}-D_{A_2}}{(a'_1 b_2+a_1 b'_2)(a'_2 b_1 +a_2 b'_1)}\leq 0,
\end{equation*}
where $D_{A_1}=a'_1 a_2 b_1 b_2-a_1 a'_2 b_1 b_2-a_1 a_2 b'_1 b_2 +a_1 a_2 b_1 b'_2+a'_1 a'_2 b_1 b'_2$, and $D_{A_2}=a'_1 a'_2 b'_1 b_2 -a'_1 a_2 b'_1 b'_2 +a_1 a'_2 b'_1 b'_2$. With this condition, we can do the following calculation
\begin{eqnarray}
  & & f_{11}^{(\alpha=A)}(m,n)-f_{11}^{(123)}(m,n) \nonumber \\
  &=&-\frac{(a_1 a'_m-a'_1 a_m)(a_1 b_n D_{A_1}+D_{A_3})}{a_1 b_1 (a_1^2+{a'_1}^2)(a_1 a'_2-a'_1 a_2)(b_1 b'_2-b'_1 b_2)} \nonumber  \\
  &\geq& -\frac{(a_1 a'_m-a'_1 a_m)(a_1 b_n D_{A_2}+D_{A_3})}{a_1 b_1 (a_1^2+{a'_1}^2)(a_1 a'_2-a'_1 a_2)(b_1 b'_2-b'_1 b_2)} \nonumber \\
  &=& \frac{(a_1 b'_2+a'_1 b_2)(a_1 a'_m-a'_1 a_m)(b_1 b'_n-b'_1 b_n)}{a_1 b_1 (a_1^2+{a'_1}^2)(b_1 b'_2-b'_1 b_2)} \nonumber \\
  &\geq& 0 \label{f11A123}
\end{eqnarray}
On the other hand, in the situation with $A\geq B$, we have
\begin{equation*}
  A-B=\frac{D_{B_1}-D_{B_2}}{(a'_1 b_2+a_1 b'_2)(a'_2 b_1 +a_2 b'_1)}\geq 0,
\end{equation*}
where $D_{B_1}=a'_1 a_2 b_1 b_2-a_1 a'_2 b_1 b_2-a_1 a_2 b'_1 b_2 +a_1 a_2 b_1 b'_2-a_1 a'_2 b'_1 b'_2$, and $D_{A_2}=a'_1 a'_2 b'_1 b_2 -a'_1 a'_2 b_1 b'_2 -a'_1 a_2 b'_1 b'_2$. With this condition, we can do the following calculation
\begin{eqnarray}
  & & f_{11}^{(\alpha=B)}(m,n)-f_{11}^{(123)}(m,n) \nonumber \\
  &=&\frac{(b_1 b'_n-b'_1 b_n)(b_1 a_m D_{B_1}+D_{B_3})}{a_1 b_1 (b_1^2+{b'_1}^2)(a_1 a'_2-a'_1 a_2)(b_1 b'_2-b'_1 b_2)} \nonumber \\
  &\geq& \frac{(b_1 b'_n-b'_1 b_n)(b_1 a_m D_{B_2}+D_{B_3})}{a_1 b_1 (b_1^2+{b'_1}^2)(a_1 a'_2-a'_1 a_2)(b_1 b'_2-b'_1 b_2)} \nonumber \\
  &=& \frac{(a_2 b'_1+a'_2 b_1)(a_1 a'_m-a'_1 a_m)(b_1 b'_n-b'_1 b_n)}{a_1 b_1 (b_1^2+{b'_1}^2)(a_1 a'_2-a'_1 a_2)} \nonumber \\
  &\geq& 0 \label{f11B123}
\end{eqnarray}
Summing up the results presented in Eq.(\ref{f11A123}) and Eq.(\ref{f11B123}), we complete the proof of Eq.(\ref{s11of123withalpha}).

In order to estimate the final key rate, we also need the upper bound of error rate caused by the two single-photon pulses, say $e_{11}$. Similar to the total gain, the total error rate with source $\alpha\beta$ chosen by Alice and Bob can be written as~\cite{ind2}
\begin{eqnarray}
  \tilde{T}_{xx}= a_1 b_1 t_{11}+a_1 b_2 t_{12} +a_2 b_1 t_{21} +a_2 b_2 t_{22}+K_{xx},&& \label{Txx} \\
  \tilde{T}_{xy}= a_1 b'_1 t_{11}+a_1 b'_2 t_{12} +a_2 b'_1 t_{21} +a_2 b'_2 t_{22}+K_{xy},&& \label{Txy} \\
  \tilde{T}_{yx}= a'_1 b_1 t_{11}+a'_1 b_2 t_{12} +a'_2 b_1 t_{21} +a'_2 b_2 t_{22}+K_{yx},&& \label{Tyx} \\
  \tilde{T}_{yy}= a'_1 b'_1 t_{11}+a'_1 b'_2 t_{12} +a'_2 b'_1 t_{21} +a'_2 b'_2 t_{22}+K_{yy},&& \label{Tyy}
\end{eqnarray}
where $T_{\alpha\beta}=E_{\alpha\beta}S_{\alpha\beta}$, $t_{mn}=s_{mn}e_{mn}$,
\begin{eqnarray}
  \tilde{T}_{xx}&=& T_{xx}-a_0 T_{0x}-b_0 T_{x0}+a_0 b_0 T_{00}, \label{tTxx} \\
  \tilde{T}_{xy}&=& T_{xy}-a_0 T_{0y}-b'_0 T_{x0}+a_0 b'_0 T_{00}, \label{tTxy} \\
  \tilde{T}_{yx}&=& T_{yx}-a'_0 T_{0x}-b_0 T_{y0}+a'_0 b_0 T_{00}, \label{tTyx} \\
  \tilde{T}_{yy}&=& T_{yy}-a'_0 T_{0y}-b'_0 T_{y0}+a'_0 b'_0 T_{00}, \label{tTyy}
\end{eqnarray}
and
\begin{eqnarray}
  K_{xx}= \sum_{(m,n)\in J_0}a_m b_n t_{mn}, & &
  K_{xy}= \sum_{(m,n)\in J_0}a_m b'_n t_{mn}, \nonumber \\
  K_{yx}= \sum_{(m,n)\in J_0}a'_m b_n t_{mn},& &
  K_{yy}= \sum_{(m,n)\in J_0}a'_m b'_n t_{mn}, \nonumber
\end{eqnarray}
with $J_0=\{(m,n)|m\geq 1,n\geq 1,m+n\geq 4,(m,n)\neq(2,2)\}$. According to Eq.(\ref{Txx}), we can find out the upper bound of $e_{11}$ such that
\begin{equation}\label{e11U}
  e_{11}\leq e_{11}^{(1)}=\frac{\tilde{T}_{xx}}{a_1 b_1 s_{11}}=\overline{e_{11}}.
\end{equation}

In the protocol, there are two different bases. We denote $s_{11}^{Z}$ and $s_{11}^{X}$ for yields of single-photon pulse pairs in the $Z$ and $X$ bases, respectively. Consider those post-selected bits cased by source $x_A x_B$ in the $Z$ basis. After an error test, we know the bit-flip error rate of this set, say $T_{x_A x_B}^{Z}=E_{x_A x_B}^{Z}S_{x_A x_B}^{Z}$. We also need the phase-flip rate for the subset of bits which are caused by the two single-photon pulse, say $e_{11}^{ph}$, which is equal to the flip rate of post-selected bits caused by a single photon in the $X$ basis, say $e_{11}^{X}$. Given this, we can now calculate the key rate by the well-know formula. For example, for those post-selected bits caused by source $y_{A} y_{B}$, it is
\begin{equation}\label{KeyRate}
  R=a'_1 b'_1 s_{11}^{Z}[1-H(e_{11}^{X})]-S_{y_A y_B}^{Z}fH(E_{y_A y_B}^{Z}),
\end{equation}
where $f$ is the efficiency factor of the error correction method used.

\section{Exact minimum of yield with only 3 states for MDI-QKD}
In the previous section, we show the lower bound of yield $s_{11}$ and the upper bound of error rate $e_{11}$ with explicit formulas. The lower bound $s_{11}^{(123)}$ is obtained with Eqs.(\ref{xx}-\ref{yx}) by setting $s_{mn}=0$, where $(m,n)\in J_1$. The upper bound of $e_{11}^{(1)}$ is obtained with Eq.(\ref{Txx}) by setting $e_{mn}=0$, where $(m,n)\in J_2$. Obviously, the condition with source $y_A y_B$ does not used in deriving $s_{11}^{(123)}$ and $e_{11}^{(1)}$. Keeping sight of this fact, we suspect that a more tightly bound can be found out with considering all relations given by Eqs.(\ref{xx}-\ref{yy}) (or Eqs.(\ref{Txx}-\ref{Tyy})). In the rest of this section, we will present an explicit algorithm within a finite number of steps to get an exact minimum of yield $s_{11}$. An an exact maximum of error rate $e_{11}$ will be given in the next section.

According to Eqs.(\ref{xx}-\ref{yy}), we can find out the expression of $s_{11},s_{12},s_{21}$ and $s_{22}$ uniquely
\begin{eqnarray}
  s_{11}&=&s_{11}^{(1234)}+\sum_{(m,n)\in J_0}f_{11}^{(1234)}(m,n)s_{mn}, \label{s11w1234} \\
  s_{12}&=&s_{12}^{(1234)}+\sum_{(m,n)\in J_0}f_{12}^{(1234)}(m,n)s_{mn}, \label{s12w1234} \\
  s_{21}&=&s_{21}^{(1234)}+\sum_{(m,n)\in J_0}f_{21}^{(1234)}(m,n)s_{mn}, \label{s21w1234} \\
  s_{22}&=&s_{22}^{(1234)}+\sum_{(m,n)\in J_0}f_{22}^{(1234)}(m,n)s_{mn}, \label{s22w1234}
\end{eqnarray}
where
\begin{eqnarray*}
  s_{11}^{(1234)}&=&\frac{a'_2 b'_2 \tilde{S}_{xx}-a'_2 b_2 \tilde{S}_{xy}-a_2 b'_2 \tilde{S}_{yx} +a_2 b_2\tilde{S}_{yy}}{(a_1 a'_2-a'_1 a_2)(b_1 b'_2-b'_1 b_2)}, \label{s11of1234} \\
  s_{12}^{(1234)}&=&\frac{-a'_2 b'_1 \tilde{S}_{xx}+a'_2 b_1 \tilde{S}_{xy}+a_2 b'_1 \tilde{S}_{yx} -a_2 b_1\tilde{S}_{yy}}{(a_1 a'_2-a'_1 a_2)(b_1 b'_2-b'_1 b_2)}, \label{s12of1234} \\
  s_{21}^{(1234)}&=&\frac{-a'_1 b'_2 \tilde{S}_{xx}+a'_1 b_2 \tilde{S}_{xy}+a_1 b'_2 \tilde{S}_{yx} -a_1 b_2\tilde{S}_{yy}}{(a_1 a'_2-a'_1 a_2)(b_1 b'_2-b'_1 b_2)}, \label{s21of1234} \\
  s_{22}^{(1234)}&=&\frac{a'_1 b'_1 \tilde{S}_{xx}-a'_1 b_1 \tilde{S}_{xy}-a_1 b'_1 \tilde{S}_{yx} +a_1 b_1\tilde{S}_{yy}}{(a_1 a'_2-a'_1 a_2)(b_1 b'_2-b'_1 b_2)}, \label{s22of1234}
\end{eqnarray*}
and
\begin{eqnarray}
  f_{11}^{(1234)}(m,n)&=&-\frac{(a_2 a'_m-a'_2 a_m)(b_2 b'_n-b'_2 b_n)}{(a_1 a'_2-a'_1 a_2)(b_1 b'_2 -b'_1 b_2)}, \label{f11of1234} \\
  f_{12}^{(1234)}(m,n)&=&-\frac{(a_2 a'_m-a'_2 a_m)(b_1 b'_n-b'_1 b_n)}{(a_1 a'_2-a'_1 a_2)(b_1 b'_2 -b'_1 b_2)}, \label{f12of1234} \\
  f_{21}^{(1234)}(m,n)&=&-\frac{(a_1 a'_m-a'_1 a_m)(b_2 b'_n-b'_2 b_n)}{(a_1 a'_2-a'_1 a_2)(b_1 b'_2 -b'_1 b_2)}, \label{f21of1234} \\
  f_{22}^{(1234)}(m,n)&=&-\frac{(a_1 a'_m-a'_1 a_m)(b_1 b'_n-b'_1 b_n)}{(a_1 a'_2-a'_1 a_2)(b_1 b'_2 -b'_1 b_2)}. \label{f22of1234}
\end{eqnarray}
In the following, we denote the superscript ${(1234)}$ by ${(*)}$ for short.

In order to estimate the lower bound of $s_{11}$, we need present some properties about the functions $f_{11}^{(*)}(m,n),f_{12}^{(*)}(m,n),f_{21}^{(*)}(m,n),f_{22}^{(*)}(m,n)$. With the condition given by Eq.(\ref{cond1}), we know that $f_{11}^{(*)}(2,k)=f_{11}^{(*)}(k,2)=0$ for all $k\geq 3$, $f_{11}^{(*)}(1,k)\geq 0, f_{11}^{(*)}(k,1)\geq 0$ for all $k\geq 3$ and $f_{11}^{(*)}(m,n)\leq 0$ for all $m,n\geq3$.
Similarly, we know that $f_{12}^{(*)}(1,k)\leq 0$, $ f_{12}^{(*)}(k,1)=f_{12}^{(*)}(2,k)=0$, $f_{12}^{(*)}(k,2)\geq 0 $, $f_{12}^{(*)}(m,n)\geq 0$, $f_{21}^{(*)}(k,1)\leq 0$, $ f_{21}^{(*)}(1,k)=f_{21}^{(*)}(k,2)=0$, $f_{21}^{(*)}(2,k)\geq 0 $, $f_{21}^{(*)}(m,n)\geq 0$, $f_{22}^{(*)}(1,k)=f_{22}^{(*)}(k,1)=0$ for all $k,m,n\geq 3$ and $f_{22}^{(*)}(m,n)\leq 0$ for all $m,n\geq 2$. With these facts, we can find out a lower bound of $s_{11}$ by setting $s_{1k}=s_{k1}=0,(k\geq 3)$ and $s_{mn}=1,(m,n\geq 3)$ crudely. Actually, all of $s_{mn},(m,n\geq 3)$ do not have to equal to 1 at the same time as the constraint conditions such that $s_{12},s_{21},s_{22}\in [0,1]$. So the problem of estimating the lower bound of $s_{11}$ can be written into the following constrained optimization problem (COP)
\begin{equation}\label{COPs11L}
  \left\{\begin{array}{cl}
    \textrm{min:} & s_{11}=s_{11}^{(*)}+\sum\limits_{(m,n)\in J_3}f_{11}^{(*)}(m,n)s_{mn} \\
    \textrm{st:} & s_{22}=s_{22}^{(*)}+\sum\limits_{(m,n)\in J_3}f_{22}^{(*)}(m,n)s_{mn}\geq 0,
  \end{array}
  \right.
\end{equation}
where $J_3=\{(m,n)|m\geq 2,n\geq 2,(m,n)\neq (2,2)\}$. In this COP problem, there are infinite number of variables. If $s_{22}^{(*)}+ \sum_{(m,n)\in J_3}f_{22}^{(*)}\geq 0$, the problem can be solved by taking $s_{mn}=1$ for all $(m,n)\in J_3$. But in practice, this trivial situation never or almost never occur.

Generally, we can not solve this COP problem analytically. In what follows we will show that the problem can be solved by explicit algorithm. Still, as shown below, it can always be determined within a finite number of steps.

\subsection{Definition of the lower bound}
In order to solve this COP problem presented in Eq.(\ref{COPs11L}), we need analysis the ratio
\begin{equation}\label{ratios11L}
  h_{11}^{(22)}(m,n)=\frac{f_{11}^{(*)}(m,n)}{f_{22}^{(*)}(m,n)} =h_a(m)\cdot h_b(n),
\end{equation}
where
\begin{equation}\label{hahb}
  h_a(m)=\frac{a_2 a'_m-a'_2 a_m}{a_1 a'_m-a'_1 a_m}, \quad h_b(n)=\frac{b_2 b'_n-b'_2 b_n}{b_1 b'_n -b'_1 b_n}.
\end{equation}
Under the condition in Eq.(\ref{cond1}), we can easily prove that $h_a(k),h_b(k)$ are two nonnegative monotone increasing functions of variable $k\geq 3$. This fact tells that $s_{m+1,n}$ and $s_{m,n+1}$ have priority to be equal to 1 over $s_{mn}$ in order to minimize $s_{11}$ in Eq.(\ref{COPs11L}). Back to the COP problem, we can solve it by introducing the following three subsets $J_L,J_s=\{(m_s,n_s)\},J_U$ of $J_3$ and one positive real number $s_L\in(0,1]$ such that
\begin{eqnarray}\label{JLU}
  h_{11}^{(22)}((m,n)\in J_L)&\leq &h_{11}^{(22)}(m_s,n_s) \nonumber \\
  & \leq& h_{11}^{(22)}((m,n)\in J_U),
\end{eqnarray}
with $J_3=J_L\cup J_s\cup J_U$ and
\begin{eqnarray}
  &s_{22}^{(*)}+\sum\limits_{(m,n)\in J_U} f_{22}^{(*)}(m,n) > 0 & \label{s22L0} \\
  &s_{22}^{(*)}+\sum\limits_{(m,n)\in J_U \cup J_s} f_{22}^{(*)}(m,n) \leq 0 & \label{s22G0} \\
  &s_{22}^{(*)}+\sum\limits_{(m,n)\in J_U} f_{22}^{(*)}(m,n) + s_L f_{22}^{(*)}(m_s,n_s)= 0.& \label{s22E0}
\end{eqnarray}
Then we can define the lower bound of $s_{11}$ by
\begin{equation}\label{s11of1234L}
  \underline{s_{11}^{*}}=s_{11}^{(*)}+ \sum_{(m,n)\in J_U} f_{11}^{(*)}(m,n) + s_L f_{11}^{(*)}(m_s,n_s).
\end{equation}
With the definitions of $J_L,J_s,J_U$ given by Eqs.(\ref{JLU}-\ref{s22E0}), we know that the set $J_{3}$ is decomposed into three subsets $J_L,J_s,J_U$. It's important to point out that the subsets $J_L,J_s,J_U$ need not be unique but the lower bound $\underline{s_{11}^{*}}$ given by Eq.(\ref{s11of1234L}) is always uniquely determined. If the subsets $J_L,J_s,J_U$ have two different choices which are denoted by $J_{L_1},J_{s_1},J_{U_1}$ and $J_{L_1},J_{s_1},J_{U_1}$, then we must have
\begin{eqnarray}\label{difJ}
  h_{11}^{(22)}((m,n)\in J_{L_{12}}) =h_{11}^{(22)}(m_{s_1},n_{s_1}) \nonumber \\
  =h_{11}^{(22)}(m_{s_2},n_{s_2}) =h_{11}^{(22)}((m,n)\in J_{U_{12}}),
\end{eqnarray}
and the number of elements in the two sets $J_{L_1},J_{L_2}$ are the same, the number of elements in the two sets $J_{U_1},J_{U_2}$ are also the same. In Eq.(\ref{difJ}), $J_{L_{12}}=(J_{L_1}- J_{L_2})\cup (J_{L_2}-J_{L_1})$ contains the elements that only included in $J_{L_{1}}$ or only in $J_{L_{2}}$ and $J_{U_{12}}=(J_{U_1}- J_{U_2})\cup (J_{U_2}- J_{U_1})$ contains the elements that only included in $J_{U_{1}}$ or only in $J_{U_{2}}$. Here and after in this article, we use $A-B$ to denote the set which contains the elements in $A$ but not in $B$. So we get $s_{L_1}=s_{L_2}$ and
\begin{eqnarray}\label{difs11L}
  & &\sum_{(m,n)\in J_{U_1}} f_{11}^{(*)}(m,n) + s_{L_1} f_{11}^{(*)}(m_{s_1},n_{s_1}) \nonumber \\
  &=&\sum_{(m,n)\in J_{U_2}} f_{11}^{(*)}(m,n) + s_{L_2} f_{11}^{(*)}(m_{s_1},n_{s_1}).
\end{eqnarray}
With this fact, we can conclude that the lower bound $\underline{s_{11}^{*}}$ given by Eq.(\ref{s11of1234L}) is uniquely.

\subsection{An algorithm for finding $J_L,J_s,J_U$ and $s_{L}$}
In order to confirm the value of $\underline{s_{11}^{*}}$, we need to determine the elements in sets $J_L,J_s,J_U$ and the proper value of $s_{L}$. In the following, we will present an algorithm for finding it within finite steps.

Though we have known that $h_{11}^{(22)}(m+1,n)\geq h_{11}^{(22)}(m,n)$ and $h_{11}^{(22)}(m,n+1)\geq h_{11}^{(22)}(m,n)$ for any $(m,n)\in J_3$. But we can not pick the larger one between $h_{11}^{(22)}(m+1,n)$ and $h_{11}^{(22)}(m,n+1)$ unless we preset the sources used by Alice and Bob. Fortunately, this defects does not affect us to derive the algorithm within finite steps.

In order to describe the algorithm clearly, we need to do the following preparations. Firstly, we define two limits
\begin{equation}\label{Limit}
  \hat{h}_a=\lim_{m\rightarrow \infty}h_a(m), \quad \hat{h}_b=\lim_{n\rightarrow \infty}h_b(n),
\end{equation}
where $h_a(m)$ and $h_b(n)$ are defined in Eq.(\ref{hahb}). As discussed before, we know that $h_a(k),h_b(k)$ are two nonnegative monotone increasing functions of $k\geq 3$. Furthermore, under the condition in Eq.(\ref{cond1}), we can also prove that $h_a(a'_m,a_m)$ is monotone increasing about $a'_m\in[0,1]$ and monotone decreasing about $a_m\in[0,1]$. Then we can find out a upper bound of the function $h_a$ such that $h_a(a'_m,a_m)\leq a_2/a_1$. By the same way, we also have $h_b(b'_n,b_n)\leq b_2/b_1$. The functions $h_a(k)$ is nonnegative monotone increasing function with finite upper bound which means that the limitations of it must be exist. And the same for $h_b(k)$. This complete the proof of Eq.(\ref{Limit}). Explicitly, if Alice and Bob send out coherent pulses, we have $\hat{h}_a=(a'_2-a_2)/(a'_1-a_1), \hat{h}_b=(b'_2-b_2)/(b'_1-b_1)$. Secondly, we also need the notations
\begin{eqnarray}
  F_c(m_0,n_0)&=&\sum_{m\geq m_0}f_{22}^{(*)}(m,n_0), \label{n0allm} \\
  F_r(m_0,n_0)&=&\sum_{n\geq n_0}f_{22}^{(*)}(m_0,n). \label{m0alln}
\end{eqnarray}
Considering the normalizing conditions
\begin{equation}\label{sum1}
  \sum_{k\geq 0}x_k=1, \quad (x=a,b,a',b'),
\end{equation}
we can calculate $F_c(m_0,n_0),F_r(m_0,n_0)$ by the following explicit formulas
\begin{eqnarray}
  F_c(m_0,n_0)&=&-\frac{(a_1 \bar{a}'_{m_0}-a'_1 \bar{a}_{m_0})(b_1 b'_{n_0}-b'_1 b_{n_0})}{(a_1 a'_2-a'_1 a_2)(b_1 b'_2 -b'_1 b_2)}, \label{Fcn0allm} \\
  F_r(m_0,n_0)&=&-\frac{(a_1 a'_{m_0}-a'_1 a_{m_0})(b_1 \bar{b}'_{n_0}-b'_1 \bar{b}_{n_0})}{(a_1 a'_2-a'_1 a_2)(b_1 b'_2 -b'_1 b_2)}, \label{Frm0alln}
\end{eqnarray}
where $\bar{x}_{k_0}=1-\sum_{k=0}^{k_0-1}x_k,(x=a,b,a',b')$.

Furthermore, for given $J_L,J_{s}$ and $J_{U}$ we introduce a vector $V_s$ with $l_s$ elements and two natural number $m_J,n_J$ such that
\begin{equation}
  m_J=\min_{(m,n)\in J_U}m,\quad n_J=\min_{(m,n)\in J_U}n.
\end{equation}
The number $l_s$ is defined by
\begin{equation}
  l_s=\max_{(m\geq m_J,n\geq n_J)\in J_L} n.
\end{equation}
The $k$-th element of $V_s$ can be defined by
\begin{equation}
  V_s(k)=\left\{
  \begin{array}{ll}
    \min_{(m,k)\in J_U} m,&  2\leq k< l_s, \\
    m_J, & k=l_s.
  \end{array}\right.
\end{equation}
Actually, with given the vector $V_s$, we know that $J_s=\{(m_s,n_s)\}$ can only be one element chosen from the following set
\begin{equation}\label{JsSet}
  \hat{J}_s=\{(m,n)=(V_s(k),k)|n_J\leq k\leq l_s\}.
\end{equation}
With $\hat{J}_s$, we define three sets which contain only one element in each of them as follows
\begin{eqnarray}
  K_s&=&\{(k,l)|h_{11}^{(22)}(k,l)=\min_{(m,n)\in \hat{J}_s}h_{11}^{(22)}(m,n)\}, \label{Ks} \\
  K_c&=&\{(k,l)|h_{11}^{(22)}(k,l)=\min_{(m,n)\in \hat{J}_s^{(c)}}h_{11}^{(22)}(m,n)\}, \label{Kc} \\
  K_r&=&\{(k,l)|h_{11}^{(22)}(k,l)=\min_{(m,n)\in \hat{J}_s^{(r)}}h_{11}^{(22)}(m,n)\}, \label{Kr}
\end{eqnarray}
where $\hat{J}_s$ is defined in Eq.(\ref{JsSet}), and $\hat{J}_s^{(c)}=\hat{J}_s-\{(V_s(n_J),n_J)\}$, $\hat{J}_s^{(r)}=\hat{J}_s-\{(V_s(l_s),l_s)\}$,

Finally, for given $J_{s}$ and $J_{U}$, we define
\begin{eqnarray}
  G(J_{U})&=&s_{22}^{(*)}+\sum_{(m,n)\in J_{U}} f_{22}^{(*)}(m,n), \label{Sumf22JU} \\
  G_s(J_{s},J_{U})&=&s_{22}^{(*)}+\sum_{(m,n)\in J_{s}\cup J_{U}} f_{22}^{(*)}(m,n). \label{Sumf22Js}
\end{eqnarray}

With these preparations, we are ready to present the algorithm as follows.

\noindent{\textit{Algorithm.}}
\\ \indent Step1. Initially, we have $J_L=\emptyset$, $J_{s}=\{(m,n)|h_{11}^{(*)}(m,n)=\min[h_{11}^{(*)}(2,3),h_{11}^{(*)}(3,2)]\}$, $J_{U}=J_3-J_s$, and $l_s=2$, if $J_s=\{(2,3)\}, V_s=(3,3)$, else if $J_s=\{(3,2)\}, V_s=(4,2)$ and $m_J=n_J=2$. Calculate $G_s$ using Eq.(\ref{Sumf22Js}). Actually, with $J_L=\emptyset$, $G_s$ can be calculated by the following explicit formula
\begin{equation}
  G_s=s_{22}^{(*)}-\frac{(a_1 \bar{a}'_{2}-a'_1 \bar{a}_{2})(b_1 \bar{b}'_{2}-b'_1 \bar{b}_{2})}{(a_1 a'_2-a'_1 a_2)(b_1 b'_2 -b'_1 b_2)}+1.
\end{equation}
As discussed before, we suppose $G'=G(L_3)<0$ initially. After these preparations, we initialize $G_f$ with $G_f=G(J_U)$ according to Eq.(\ref{Sumf22JU}). If $G_f<0$ we goto Step2, else we goto Step3.
\\ \indent Step2. Find out $J_s'=\{(m_s',n_s')\}$ such that $f_{22}^{(*)}(m_s',n_s')=\min_{(m,n)\in J_{U}} f_{22}^{(*)}(m,n)=\min_{(m,n)\in \hat{J}_{s}} f_{22}^{(*)}(m,n)$, where $\hat{J}_s$ is defined in Eq.(\ref{JsSet}). So we need to find out the set $\hat{F}_s$. Firstly, we check that the following two inequalities fulfilled or not,
\begin{eqnarray}
  h_{11}^{(22)}((m,n)\in K_r)&\geq& h_a(V_s(n_J))\cdot \hat{h}_b, \label{AllRaw} \\
  h_{11}^{(22)}((m,n)\in K_c)&\geq& \hat{h}_a \cdot h_b(l_s), \label{AllColumn}
\end{eqnarray}
where $h_a(k),h_b(k)$ are defined in Eq.(\ref{hahb}) and $\hat{h}_a,\hat{h}_b$ are defined in Eq.(\ref{Limit}). If IEq.(\ref{AllRaw}) holds, we goto Step2.1, else if IEq.(\ref{AllColumn}) holds, we goto Step2.2, else we goto Step2.3.
\\ \indent  Step2.1. In this situation, we know that $h_{11}^{(22)}((m,n)\in K_r)$ is greater than all the values $h_{11}^{(22)}(V_s(l_s),k)$ for all $k\geq 2$. We need to calculate the value
\begin{equation}
  G'(J_U)=G_f-\sum_{(m,n)\in J_r}f_{22}^{(*)}(m,n), \label{GpJU}
\end{equation}
where $J_r=\{(m,n)|m=V_s(l_s),n\geq l_s\}$. If $G'(J_U)\leq 0$ we need remove all the elements in $J_r$ from the set of $J_U$. Then we can renew the values with $m_J=m_J+1,n_J=n_J,l_s=l_s-1,J_L=J_L\cup J_r, J_s=K_r, J_U=J_3-J_L-J_s$ and $G_f=G_f-f_{22}^{(*)}((m,n)\in J_s)$. If $G_f< 0$, we go back to step2, else we goto Step3. On the other hand, if $G'(J_U)> 0$, then we know that the element in the final set of $J_s$ must be included in $J_r$. In this case we need to renew $J_L=J_L\cup J_s, J_s= \{(V_s(l_s),l_s)\}, J_U=J_U-J_s, V_s(l_s)=V_s(l_s)+1, l_s=l_s+1, V_s(l_s)=V_s(l_s-1)-1$, and $G_f=G_f-f_{22}^{(*)}((m,n)\in J_s)$. If $G_f<0 $, we go back to step2, else we goto Step3.
\\ \indent  Step2.2. In this situation, we know that $h_{11}^{(22)}((m,n)\in K_c)$ is greater than all the values $h_{11}^{(22)}(k,n_J)$ for all $k\geq 2$. We need to calculate the value
\begin{equation}
  G'(J_U)=G_f-\sum_{(m,n)\in J_c}f_{22}^{(*)}(m,n), \label{GpJU}
\end{equation}
where $J_c=\{(m,n)|m\geq V_s(n_J),n=n_J\}$. If $G'(J_U)\leq 0$ we need remove all the elements in $J_c$ from the set of $J_U$. Then we can renew the values with $m_J=m_J,n_J=n_J+1,l_s=l_s,J_L=J_L\cup J_c, J_s=K_c, J_U=J_3-J_L-J_s$ and $G_f=G_f-f_{22}^{(*)}((m,n)\in J_s)$. If $G_f<0 $, we go back to step2, else we goto Step3. On the other hand, if $G'(J_U)> 0$, then we know that the element in the final set of $J_s$ must be included in $J_c$. In this case we need to renew $J_L=J_L\cup J_s, J_s= \{(V_s(n_J),n_J)\}, J_U=J_U-J_s, V_s(n_J)=V_s(n_J)+1, l_s=l_s$, and $G_f=G_f-f_{22}^{(*)}((m,n)\in J_s)$. If $G_f<0 $, we go back to step2, else we goto Step3.
\\ \indent Step2.3. In this situation, we should renew $J_L=J_L\cup J_s, J_s=K_s, J_U=J_U-J_s$. Denoting $K_s=\{(k_m,k_n)\}$, we can renew $l_s$ and $V_s$ by the following method. If $k_n=l_s$, we have $V_s(l_s)=V_s(l_s)+1, l_s=l_s+1, V_s(l_s)=V_s(l_s-1)-1$. If $k_n<l_s$, we have $V_s(k_n)=V_s(k_n)+1, l_s=l_s$. Finally, we renew $G_f=G_f-f_{22}^{(*)}((m,n)\in J_s)$. If $G_f<0 $, we go back to step2, else we goto Step3.
\\ \indent Step3. Now we have already find out the final sets $J_{L},J_s,J_U$ with Step2. In this step, we will calculate the value of $s_L$. According to the relation presented in Eq.(\ref{s22E0}), we should define
\begin{equation}\label{sL}
  s_L=-\frac{G_f}{f_{22}^{(*)}((m,n)\in J_s)}.
\end{equation}
Then we can calculate the lower bound $\underline{s_{11}^{*}}$ of $s_{11}$ by using Eq.(\ref{s11of1234L}).

\section{Exact maximum error rate with only 3 states for MDI-QKD}
In section 2, we show the upper bound of error rate $e_{11}$ with explicit formula. The upper bound of $e_{11}^{(1)}$ is obtained with Eq.(\ref{Txx}) by setting $e_{mn}=0$, where $(m,n)\in J_2$. Obviously, the condition with source $y_A$ and $y_B$ does not used in deriving $e_{11}^{(1)}$. Keeping sight of this fact, we suspect that a more tightly bound can be found out with considering all relations given by Eqs.(\ref{Txx}-\ref{Tyy}). In the rest of this section, we will present that we can find out an exact maximum of error rate $e_{11}$ within a finite number steps.

According to Eqs.(\ref{Txx}-\ref{Tyy}), we can find out the expression of $t_{11},t_{12},t_{21}$ and $t_{22}$ uniquely
\begin{eqnarray}
  t_{11}&=&t_{11}^{(*)}+\sum_{(m,n)\in J_0}f_{11}^{(*)}(m,n)t_{mn}, \label{t11w1234} \\
  t_{12}&=&t_{12}^{(*)}+\sum_{(m,n)\in J_0}f_{12}^{(*)}(m,n)t_{mn}, \label{t12w1234} \\
  t_{21}&=&t_{21}^{(*)}+\sum_{(m,n)\in J_0}f_{21}^{(*)}(m,n)t_{mn}, \label{t21w1234} \\
  t_{22}&=&t_{22}^{(*)}+\sum_{(m,n)\in J_0}f_{22}^{(*)}(m,n)t_{mn}, \label{t22w1234}
\end{eqnarray}
where
\begin{eqnarray*}
  t_{11}^{(*)}&=&\frac{a'_2 b'_2 \tilde{T}_{xx}-a'_2 b_2 \tilde{T}_{xy}-a_2 b'_2 \tilde{T}_{yx} +a_2 b_2\tilde{T}_{yy}}{(a_1 a'_2-a'_1 a_2)(b_1 b'_2-b'_1 b_2)}, \label{t11of1234} \\
  t_{12}^{(*)}&=&\frac{-a'_2 b'_1 \tilde{T}_{xx}+a'_2 b_1 \tilde{T}_{xy}+a_2 b'_1 \tilde{T}_{yx} -a_2 b_1\tilde{T}_{yy}}{(a_1 a'_2-a'_1 a_2)(b_1 b'_2-b'_1 b_2)}, \label{t12of1234} \\
  t_{21}^{(*)}&=&\frac{-a'_1 b'_2 \tilde{T}_{xx}+a'_1 b_2 \tilde{T}_{xy}+a_1 b'_2 \tilde{T}_{yx} -a_1 b_2\tilde{T}_{yy}}{(a_1 a'_2-a'_1 a_2)(b_1 b'_2-b'_1 b_2)}, \label{t21of1234} \\
  t_{22}^{(*)}&=&\frac{a'_1 b'_1 \tilde{T}_{xx}-a'_1 b_1 \tilde{T}_{xy}-a_1 b'_1 \tilde{T}_{yx} +a_1 b_1\tilde{T}_{yy}}{(a_1 a'_2-a'_1 a_2)(b_1 b'_2-b'_1 b_2)}, \label{t22of1234}
\end{eqnarray*}
and $f_{11}^{(*)}(m,n),f_{12}^{(*)}(m,n),f_{21}^{(*)}(m,n),f_{22}^{(*)}(m,n)$ are defined by Eqs(\ref{f11of1234}-\ref{f22of1234}).

With the properties of the functions $f_{11}^{(*)}(m,n), f_{12}^{(*)}(m,n), f_{21}^{(*)}(m,n), f_{22}^{(*)}(m,n)$ discussed in the previous section, we can find out a upper bound of $t_{11}$ by setting $t_{1k}=t_{k1}=1,(k\geq 3)$ and $t_{mn}=0,(m,n\geq 3)$ crudely. As discussed in the previous section, all of $t_{1k},t_{k1},(k\geq 3)$ do not have to equal to 1 at the same time as the constraint conditions such that $t_{12},t_{21},t_{22}\in [0,1]$. So the problem of estimating the upper bound of $t_{11}$ can be written into the following constrained optimization problem (COP)
\begin{equation}\label{COPt11U}
  \left\{\begin{array}{cl}
    \textrm{max:} & t_{11}=t_{11}^{(*)}+\sum\limits_{k\geq 3}f_{11}^{(*)}(1,k)t_{1k} +f_{11}^{(*)}(k,1)t_{k1}, \\
    \textrm{st:} & t_{12}=t_{12}^{(*)}+\sum\limits_{k\geq 3}f_{12}^{(*)}(1,k)t_{1k}\geq 0, \\
     & t_{12}=t_{12}^{(*)}+\sum\limits_{k\geq 3}f_{21}^{(*)}(k,1)t_{k1}\geq 0,
  \end{array}
  \right.
\end{equation}
In this COP problem, there are infinite number of variables. Considering the independence between variables $t_{1k}$ and $t_{k1}$, the COP problem in Eq.(\ref{COPt11U}) can be decomposed into the following two COP problems
\begin{equation}\label{COPt11U1k}
  \left\{\begin{array}{cl}
    \textrm{max:} & t_{1}=\sum\limits_{k\geq 3}f_{11}^{(*)}(1,k)t_{1k}, \\
    \textrm{st:} & t_{12}=t_{12}^{(*)}+\sum\limits_{k\geq 3}f_{12}^{(*)}(1,k)t_{1k}\geq 0,
  \end{array}
  \right.
\end{equation}
and
\begin{equation}\label{COPt11Uk1}
  \left\{\begin{array}{cl}
    \textrm{max:} & t_{2}=\sum\limits_{k\geq 3}f_{11}^{(*)}(k,1)t_{k1}, \\
    \textrm{st:} & t_{12}=t_{12}^{(*)}+\sum\limits_{k\geq 3}f_{21}^{(*)}(k,1)t_{k1}\geq 0,
  \end{array}
  \right.
\end{equation}

In order to solve the COP problems, we need analysis the ratio $f_{11}^{(*)}(1,k)/f_{12}^{(*)}(1,k)$ and $f_{11}^{(*)}(k,1)/f_{21}^{(*)}(k,1)$. Actually, we have
\begin{equation*}\label{f111kk1}
  \frac{f_{11}^{(*)}(1,k)}{f_{12}^{(*)}(1,k)}=-h_b(b'_k,b_k),\quad \frac{f_{11}^{(*)}(k,1)}{f_{21}^{(*)}(k,1)}=-h_a(a'_k,a_k),
\end{equation*}
where $h_a,h_b$ are defined in Eq.(\ref{hahb}). As discussed before, under the condition in Eq.(\ref{cond1}), $h_a(k),h_b(k)$ are two nonnegative monotone increasing functions of variable $k\geq 3$. This fact predicts that $t_{1,k+1}$ ($t_{k+1,1}$) has priority to be equal to 1 over $t_{1,k}$ ($t_{k,1}$) in order to maximize $t_1$ ($t_2$). Back to the COP problems, we can solve them by introducing two neutral numbers $k_a,k_b$ and two positive real numbers $s_a,s_b\in(0,1]$ such that
\begin{equation}\label{kasa}
 \left\{\begin{array}{l}
  t_{12}^{(*)}+\sum_{k> k_b} f_{12}^{(*)}(1,k)> 0  \\
  t_{12}^{(*)}+\sum_{k\geq k_b} f_{12}^{(*)}(1,k)\leq 0  \\
  t_{12}^{(*)}+\sum_{k\geq k_b} f_{12}^{(*)}(1,k)+s_b f_{12}^{(*)}(1,k_b)=0,
 \end{array}\right.
\end{equation}
and
\begin{equation}\label{kbsb}
 \left\{\begin{array}{l}
  t_{21}^{(*)}+\sum_{k> k_a} f_{21}^{(*)}(k,1)> 0  \\
  t_{21}^{(*)}+\sum_{k\geq k_a} f_{21}^{(*)}(k,1)\leq 0  \\
  t_{21}^{(*)}+\sum_{k\geq k_a} f_{21}^{(*)}(k,1)+s_a f_{21}^{(*)}(k_a,1)=0.
 \end{array}\right.
\end{equation}
Then we can define the upper bound of $t_{11}$ by
\begin{eqnarray}\label{t11of1234U}
  \overline{t_{11}^{*}}&=&t_{11}^{(*)}+\sum_{k\geq k_b}f_{11}^{(*)}(1,k)+s_b f_{11}^{(*)}(1,k_b) \nonumber \\
  & &+\sum_{k\geq k_a}f_{11}^{(*)}(k,1) +s_a f_{11}^{(*)}(k_a,1),
\end{eqnarray}
where $k_a,k_b$ and $s_a,s_b$ are defined in Eqs.(\ref{kasa}-\ref{kbsb}) which can be easily find out by using the Algorithm presented in the previous section.

After getting the lower bound of $s_{11}$ and the upper bound of $t_{11}$, we can easily obtain the upper bound of the error rate $e_{11}$ such that
\begin{equation}\label{e11of1234U}
  \overline{e_{11}^{(*)}}=\overline{t_{11}^{(*)}}\bigg{/}\underline{s_{11}^{(*)}},
\end{equation}
where $\underline{s_{11}^{(*)}}$ is defined in Eq.(\ref{s11of1234L}) and $\overline{t_{11}^{(*)}}$ is defined in Eq.(\ref{t11of1234U}).

\section{Numerical Simulation}
\begin{figure}
  \includegraphics[width=240pt]{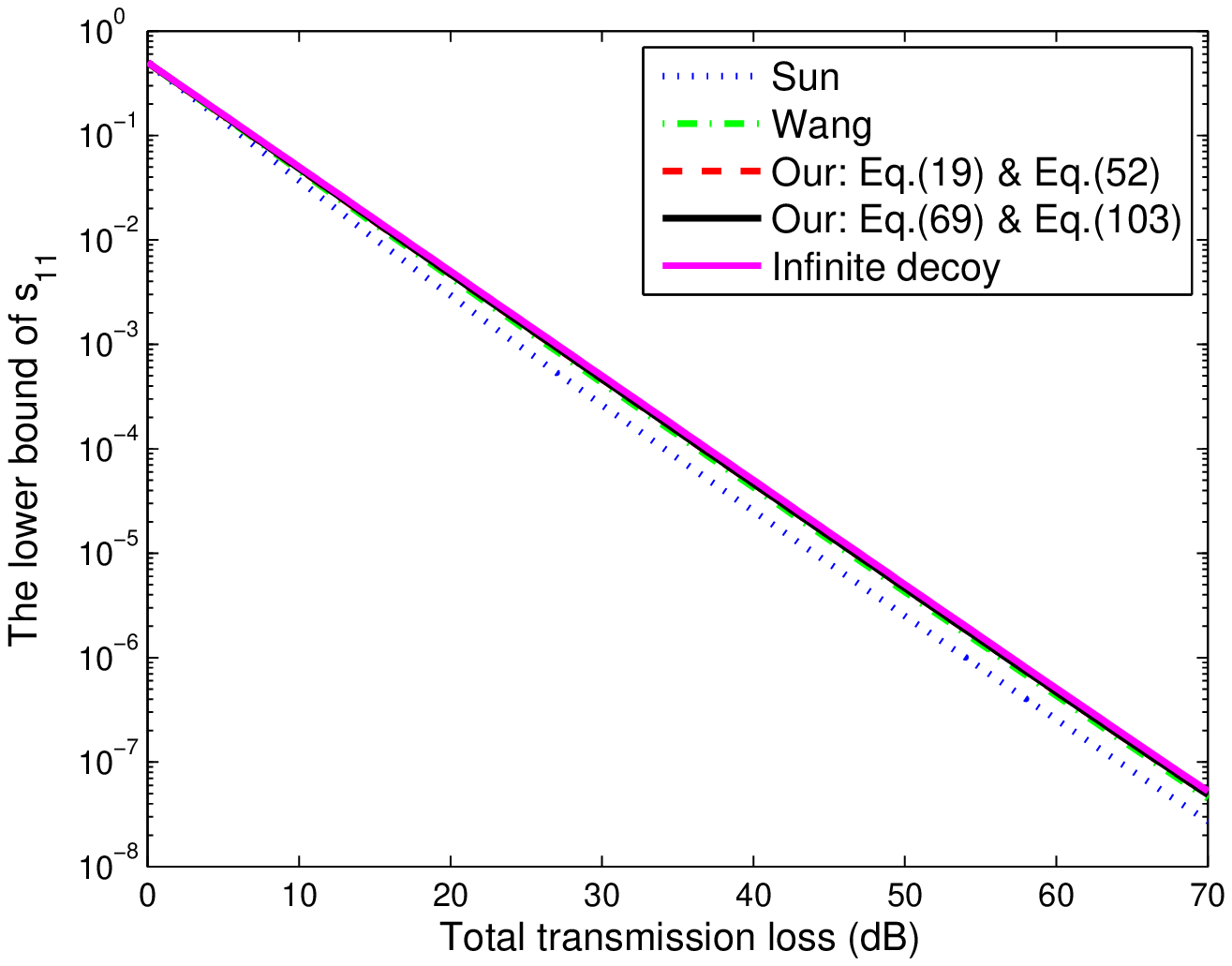}\\
  \includegraphics[width=240pt]{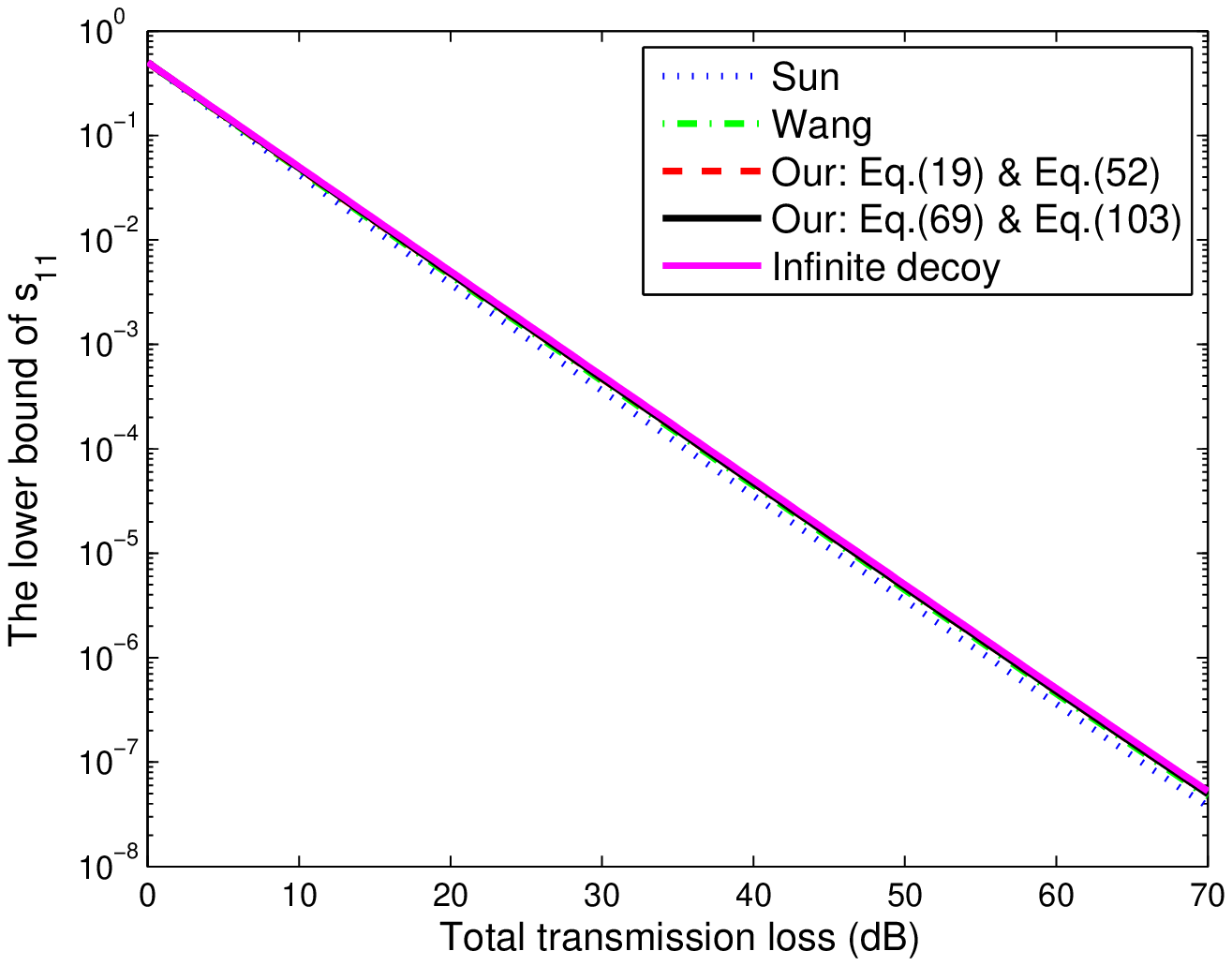}\\
  \caption{(Color online) The estimated parameter of $s_{11}$ versus the total channel transmission loss using 3-intensity decoy state MDI-QKD.}\label{s11Lp1p5}
\end{figure}

\begin{figure}
  \includegraphics[width=240pt]{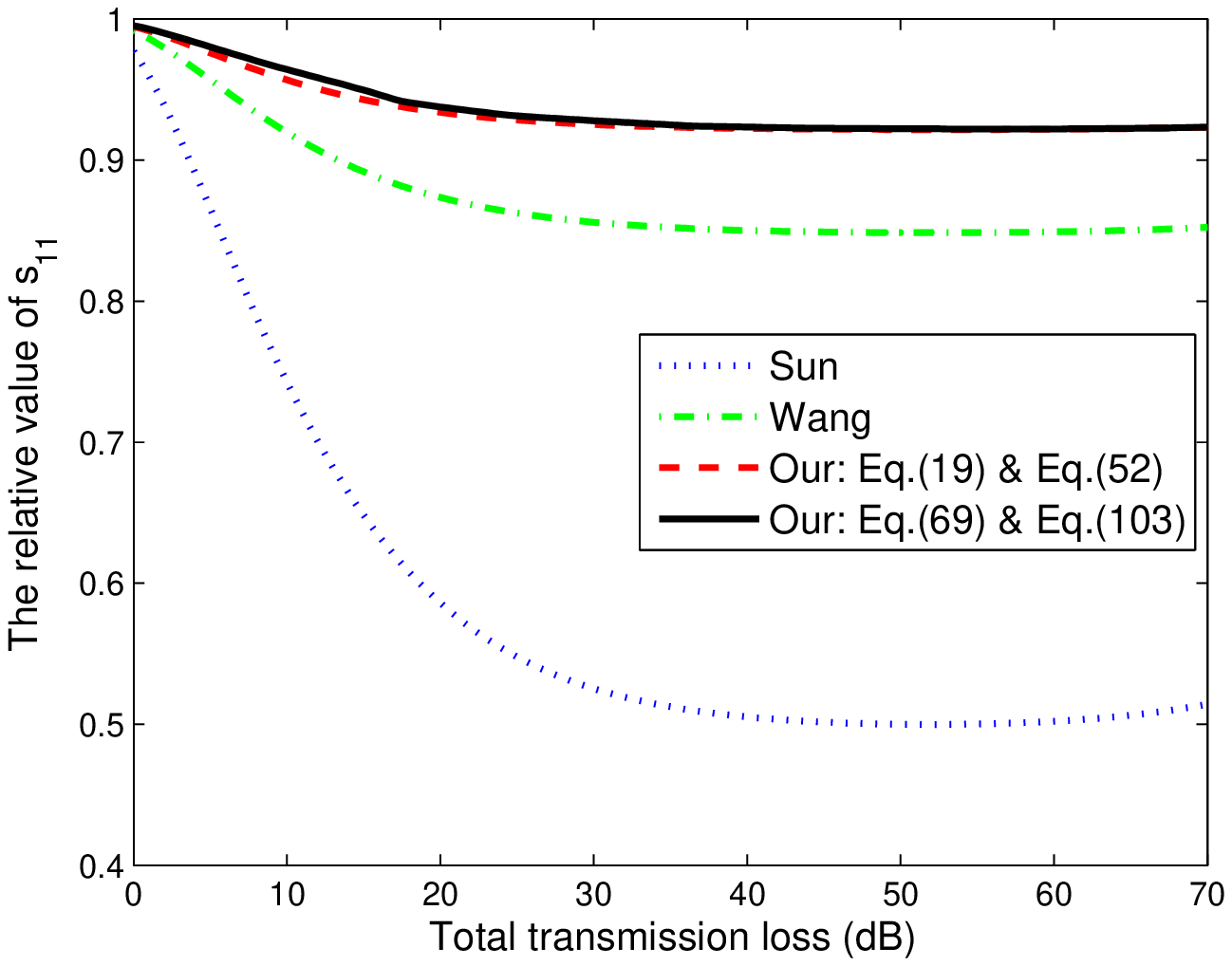}\\
  \includegraphics[width=240pt]{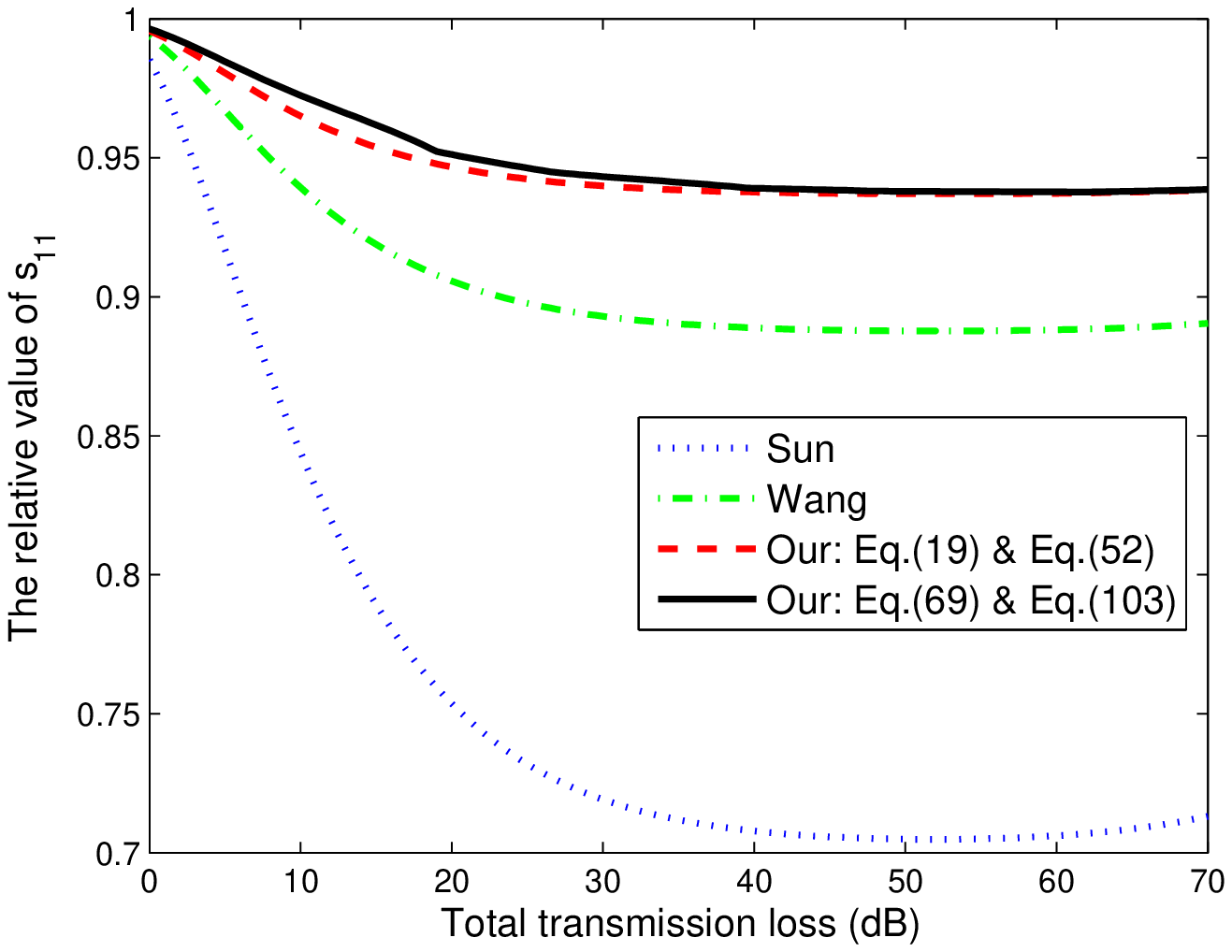}\\
  \caption{(Color online) The relative value between estimated parameter of $s_{11}$ and the asymptotic limit of the infinite decoy-state method versus the total channel transmission loss using 3-intensity decoy state MDI-QKD.}\label{rs11Lp1p5}
\end{figure}

\begin{figure}
  \includegraphics[width=240pt]{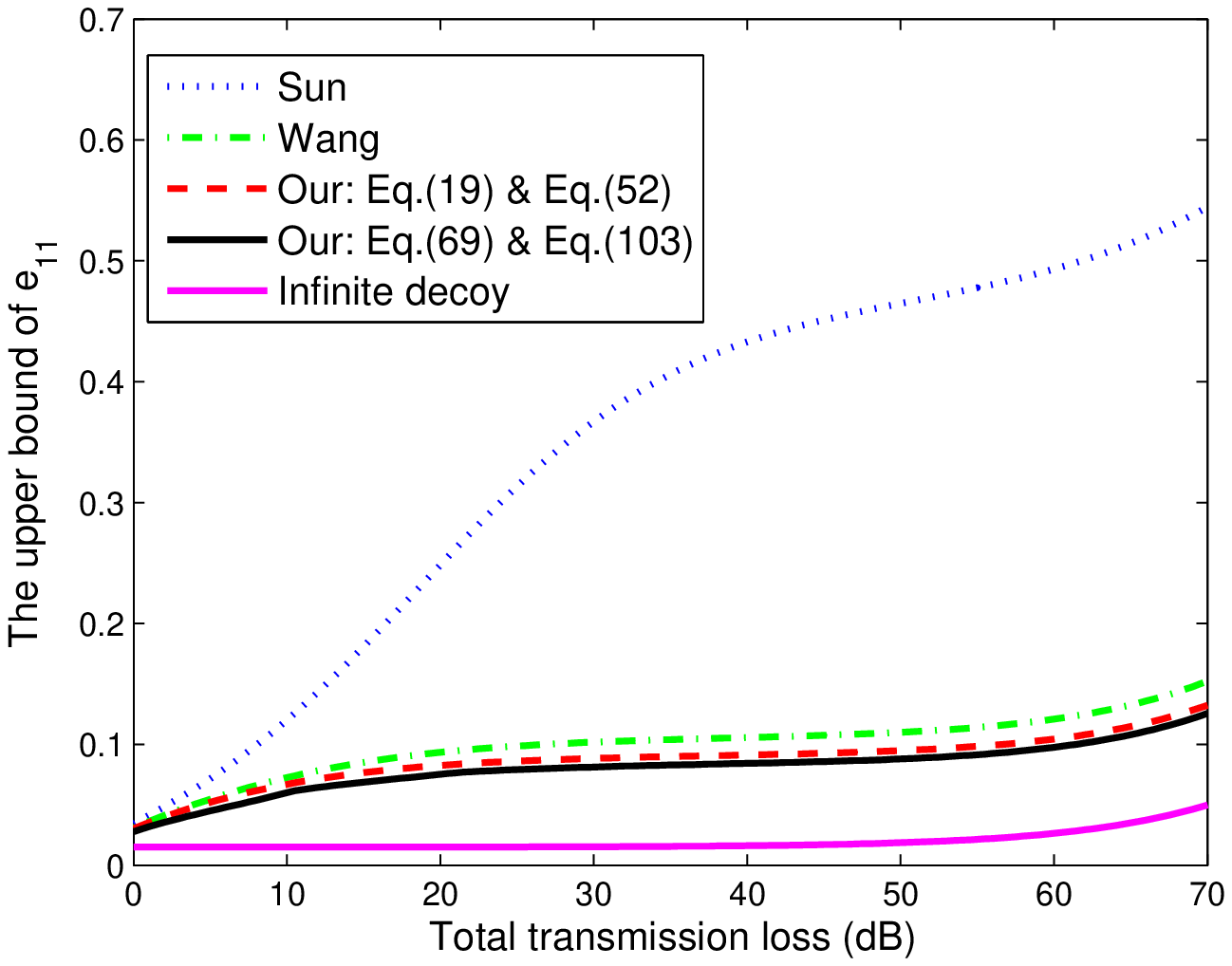}\\
  \includegraphics[width=240pt]{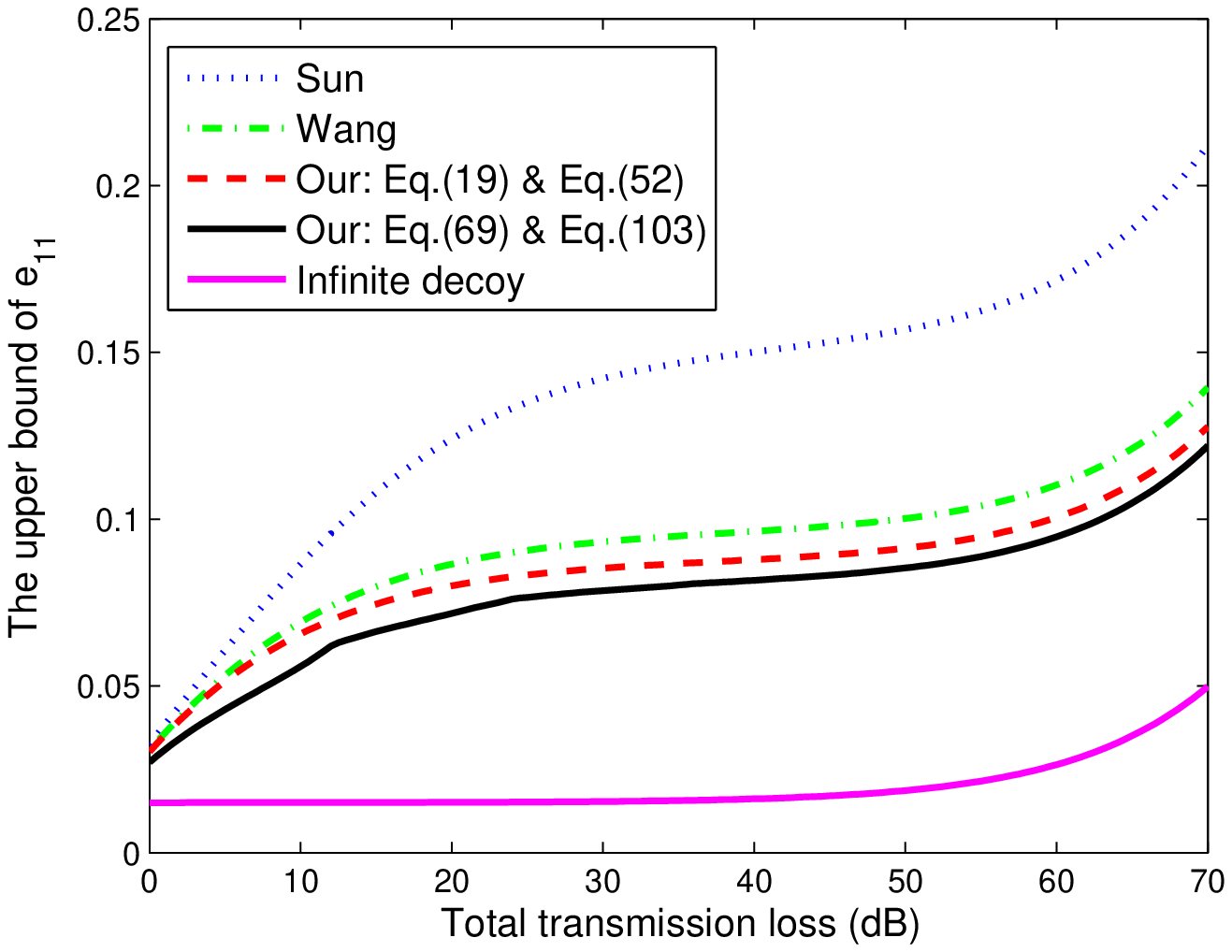}\\
  \caption{(Color online) The estimated parameter of $e_{11}$ versus the total channel transmission loss using 3-intensity decoy state MDI-QKD.}\label{e11Up1p5}
\end{figure}

\begin{figure}
  \includegraphics[width=240pt]{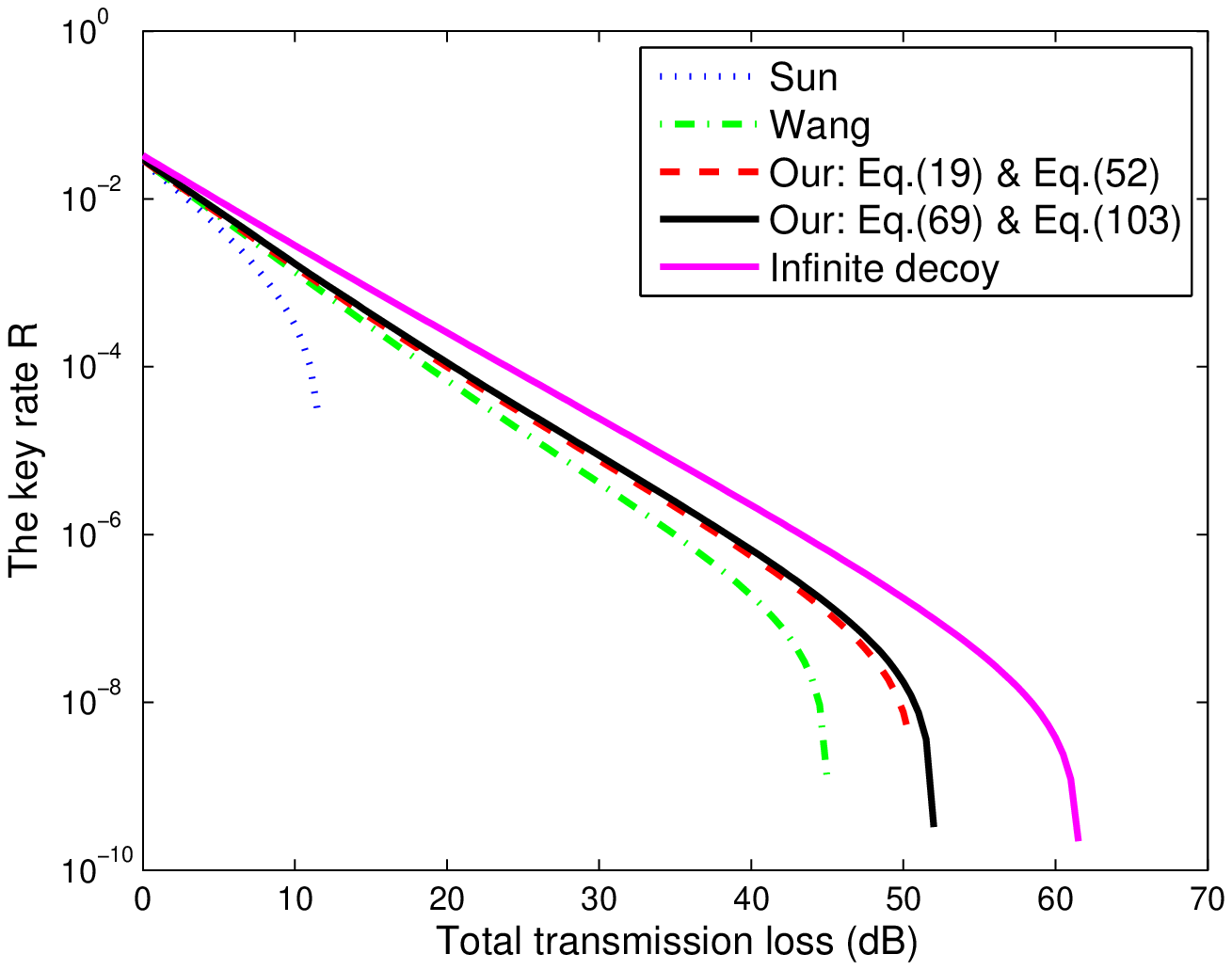}\\
  \includegraphics[width=240pt]{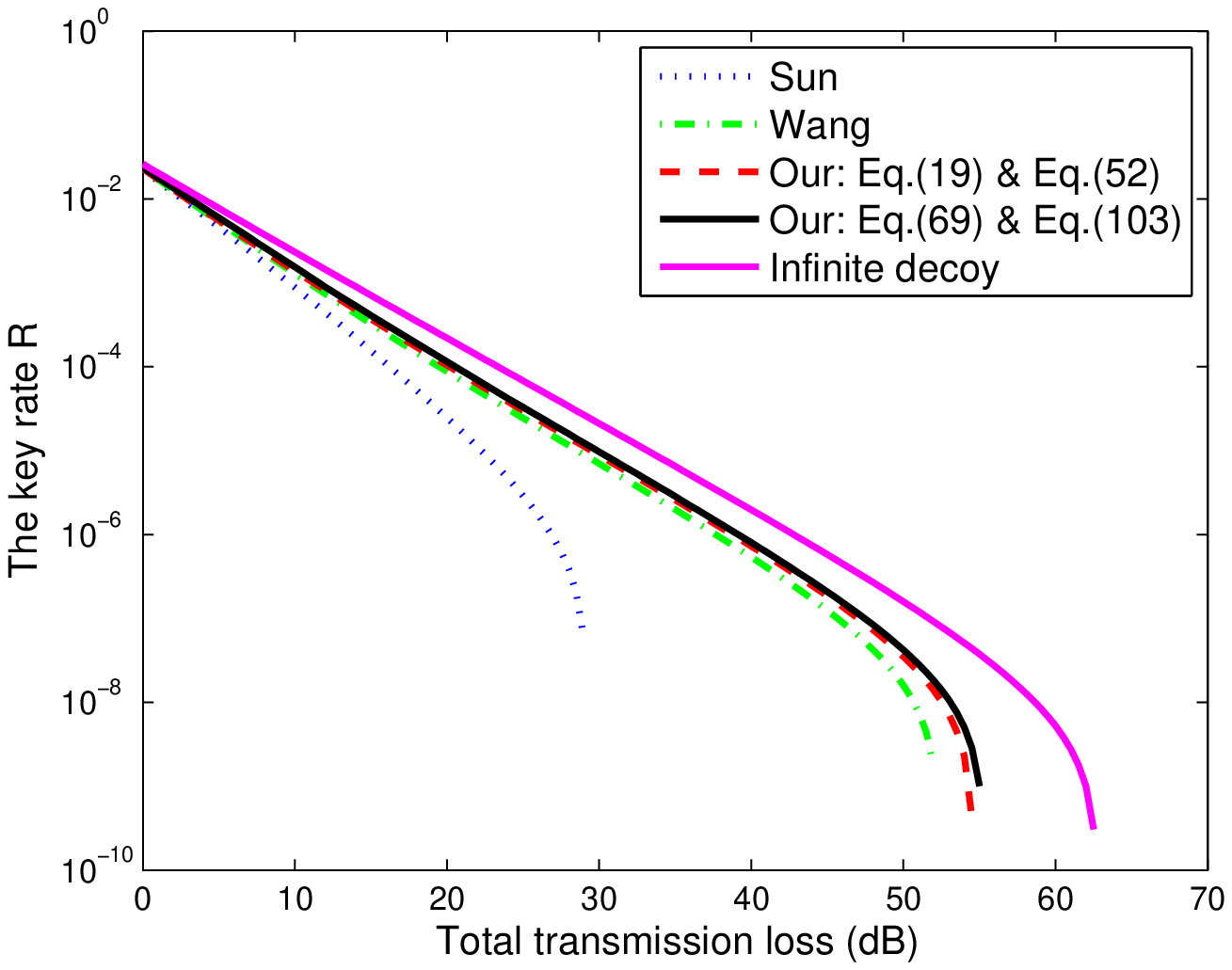}\\
  \caption{(Color online) The estimated key rate $R$ versus channel transmission using 3-intensity decoy state MDI-QKD.}\label{Rp1p5}
\end{figure}

\begin{figure}
  \includegraphics[width=240pt]{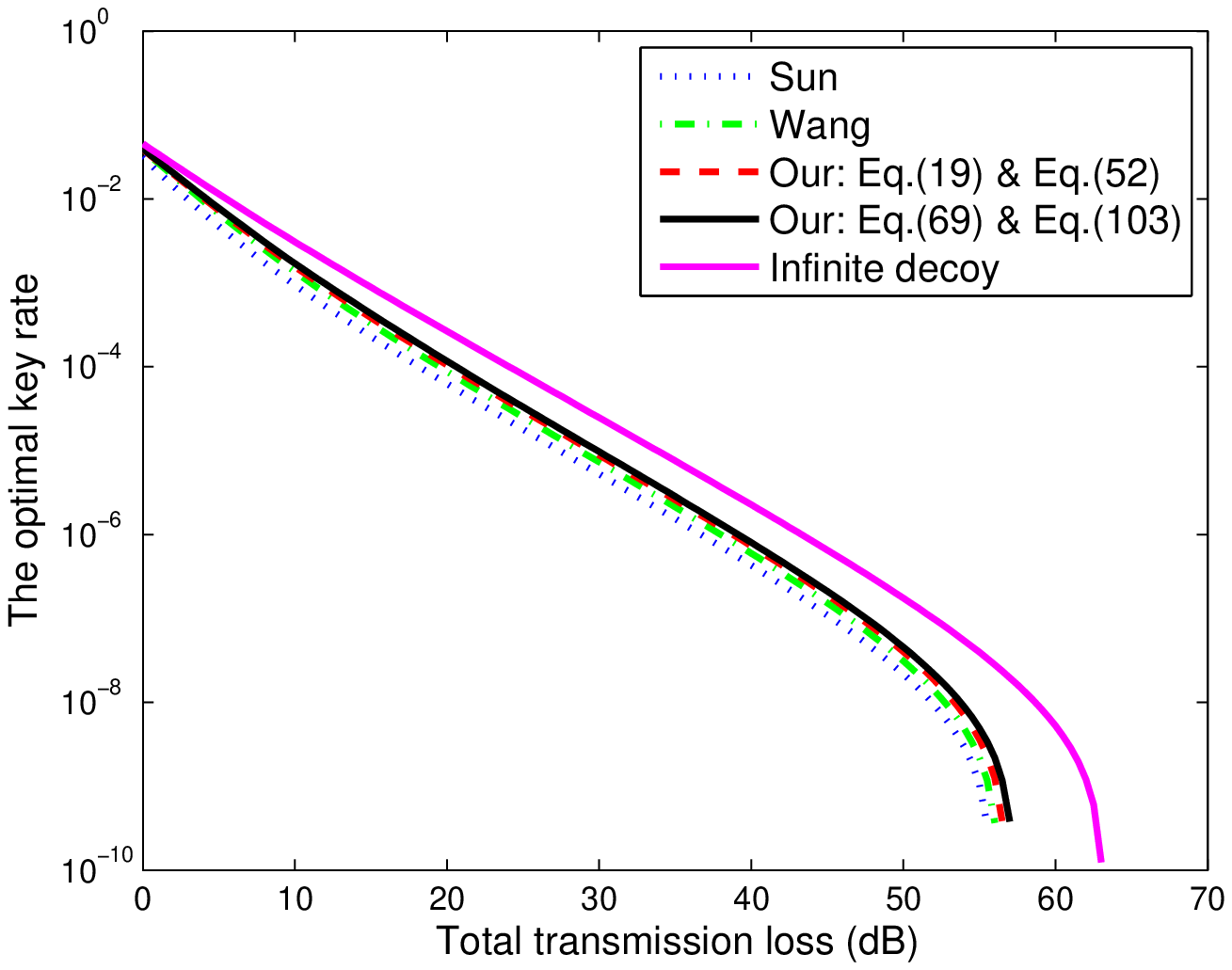}\\
  \caption{(Color online) The optimal key rate versus the total channel transmission loss using 3-intensity decoy state MDI-QKD.}\label{OptR}
\end{figure}

\begin{figure}
  \includegraphics[width=240pt]{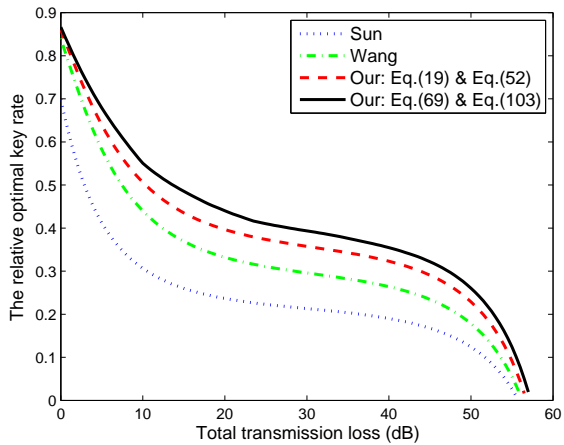}\\
  \caption{(Color online) The relative value between the optimal key rate obtained with different methods and the asymptotic limit of the infinite decoy-state method versus the total channel transmission loss using 3-intensity decoy state MDI-QKD.}\label{rOptR}
\end{figure}

\begin{figure}
  \includegraphics[width=240pt]{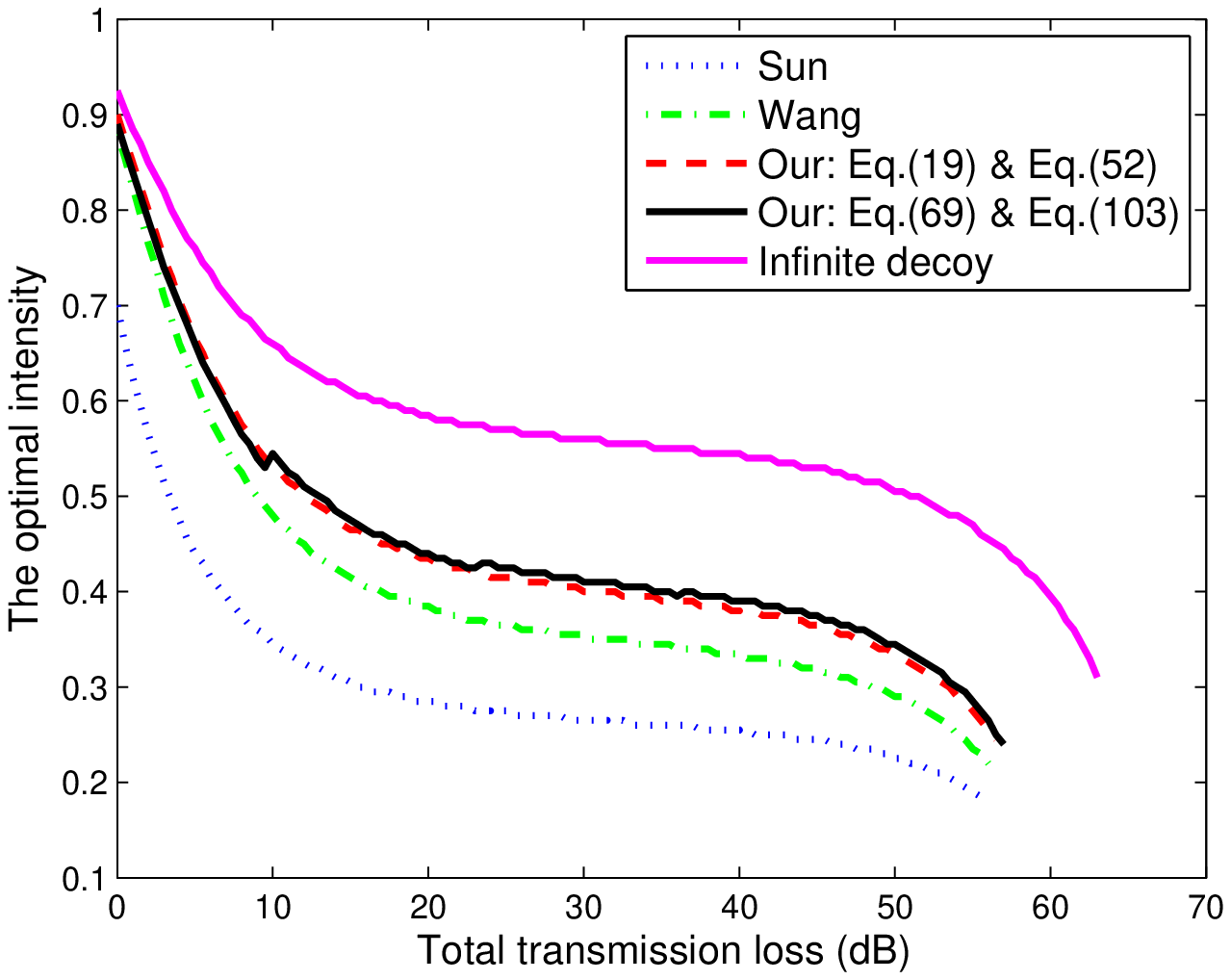}\\
  \caption{(Color online) The optimal key rate versus the total channel transmission loss using 3-intensity decoy state MDI-QKD.}\label{OptRmu}
\end{figure}

\begin{table}
\caption{\label{tabPara}List of experimental parameters used in numerical simulations: $e_0$ is the error rate of background, $e_d$ is the misalignment-error probability; $p_d$ is the dark count rate per detector; $f$ is the error correction inefficiency.}
\begin{ruledtabular}
\begin{tabular}{cccc}
  $e_0$ & $e_d$ & $p_d$ & $f$ \\
  \hline
  0.5 & 1.5\% & $3.0\times 10^{-6}$ & 1.16\\
\end{tabular}
\end{ruledtabular}
\end{table}

In this section, we will present some numerical simulations to comparing our results with the pre-existing results~\cite{wangPRA2013,LiangPRA2013}. Below for simplicity, we suppose that Alice and Bob use the coherent sources. The UTP locates in the middle of Alice and Bob, and the UTP's detectors are identical, i.e., they have the same dark count rate and detection efficiency, and their detection efficiency does not depend on the incoming signals. We shall estimate what values would be probably observed for the gains and error rates in the normal cases by the linear models as in~\cite{wang05,ind2,LiangPRA2013}:
\begin{eqnarray*}
  |n\rangle\langle n| = \sum_{k=0}^n C_n^k \xi^k (1-\xi)^{n-k}|k\rangle\langle k|
\end{eqnarray*}
where $\xi^k$ is the transmittance for a distance from Alice to the UTB.  For fair comparison, we use the same parameter values used in~\cite{ind2,LiangPRA2013} for our numerical evaluation, which follow the experiment reported in~\cite{UrsinNP2007}. For simplicity, we shall put the detection efficiency to the overall transmittance $\eta=\xi^2 \zeta$. We assume all detectors have the same detection efficiency $\zeta$ and dark count rate $p_d$. The values of these parameters are presented in Table~\ref{tabPara}. With this, the total gains $S_{\mu_i,\nu_j}^{\omega},(\omega=X,Z)$ and error rates $S_{\mu_i,\nu_j}^{\omega}E_{\mu_i,\nu_j}^{\omega},(\omega=X,Z)$ of Alice's intensity $\mu_i (i=0,1,2)$ and Bob's intensity $\nu_j (j=0,1,2)$ can be calculated. By using these values, we can estimate the lower bounds of yield $s_{11}^{Z}$ with Eq.(\ref{s11L123}) and Eq.(\ref{s11of1234L}). Also, we can estimate the upper bounds of error rate $e_{11}^{X}$ with Eq.(\ref{e11U}) and Eq.(\ref{e11of1234U}). The estimated parameters of $s_{11}$ and $e_{11}$ are shown in Fig.\ref{s11Lp1p5} and Fig.\ref{e11Up1p5}, respectively, which clearly shows that our methods are more tightly than the pre-existed results. In order to see more clearly, in Fig.\ref{rs11Lp1p5}, we plot the relative value of $s_{11}$ to the result obtained with the infinite decoy-state method. We can observe that our results are more close to the asymptotic limit of the infinite decoy-state method than the pre-existed results. Moreover, we can find out that the results given by the numerical method are more tightly than the analytical method. Furthermore, with these parameters, we can estimate the final key rate $R$ of this protocol with Eq.(\ref{KeyRate}) which is shown in Fig.\ref{Rp1p5}. In these four figures, the blue dotted line is obtained by the method presented in Ref.~\cite{LiangPRA2013}, the green dash-dot line is obtained by the method presented in Ref.~\cite{wangPRA2013}, the red dashed line is obtained by the analytical method presented in section 2 with Eq.(\ref{s11L123}) and Eq.(\ref{e11U}), the black solid line is obtained by the numerical method presented in section 3 with Eq.(\ref{s11of1234L}) and Eq.(\ref{e11of1234U}) and the magenta solid line is the result obtained by the infinite decoy-state method. In the simulation, the densities used by Alice and Bob are assigned to $\mu_1=\nu_1=0.1$, $\mu_2=\nu_2=0.5$.

Furthermore, if we fix the densities of the decoy-state pulses used by Alice and Bob, the final key rate will change with Alice and Bob taking different densities for their single-state pulses. Here, we also take $\mu_1=\nu_1=0.1$ and assume that $\mu_2=\nu_2>\mu_1$. In Fig.\ref{OptR}, we present the the optimal key rates with different methods. In order to see more clearly, in Fig.\ref{rOptR}, we plot the relative value of the optimal key rate to the result obtained with the infinite decoy-state method. We can observe that our results are better than the pre-existed results. The optimal densities with the optimal key rate versus the total channel transmission loss is given in Fig.\ref{OptRmu}.

\section{Conclusion}
We study the MDI-QKD in practice with only 3 different states in implementing the decoy-state method. Firstly, we present a more tightly analytical formulas for the decoy-state method for two-pulse sources with 3 different states. Then we show an exact maximum of the yield $s_{11}$ and an exact minimum of the error rate $e_{11}$ with an efficient algorithm. These methods can be applied to the recently proposed MDI-QKD with imperfect single-photon source such as the coherent states or the heralded states from the parametric down conversion. Finally, we give some numerical simulations. The results show that our methods are better than the pre-existing results.

{\bf Acknowledgement:}
We acknowledge
the support from the 10000-Plan of Shandong province,
the National High-Tech Program of China Grants
No. 2011AA010800 and No. 2011AA010803 and NSFC
Grants No. 11174177 and No. 60725416.


\end{document}